%% file: main.tex
\documentclass[fleqn,usenatbib,useAMS]{mnras}

\usepackage{graphicx}	
\usepackage{amsmath}	
\usepackage{amssymb}	
\usepackage{multicol}        
\usepackage{bm}		
\usepackage{pdflscape}	
\usepackage{float}

\usepackage[flushleft]{threeparttable}
\usepackage{amsmath,empheq}
\usepackage{makecell}
\usepackage{orcidlink}
\newcommand{\orcid}{\orcidlink}
\usepackage[T1]{fontenc}
\usepackage{ae,aecompl}

\usepackage{newtxtext,newtxmath}

\usepackage{cases}
\usepackage{graphicx}
\usepackage{CJK}
\usepackage{subfigure}
\usepackage{natbib}
\usepackage{color}
\usepackage{tabularx}
\usepackage{bm}
\usepackage{threeparttable}

\newcommand{\thetae}{\theta_{\rm E}}

\newcommand{\te}{t_{\rm E}}

\usepackage{calc}
\usepackage{xcolor, xspace}
\definecolor{mycolor0}{rgb}{0,0,0}
\definecolor{mycolor1}{rgb}{0,0,0}
\definecolor{cRsp1}{rgb}{0,0,0} 
\definecolor{cRsp2}{rgb}{0,0,0}

\newcommand{\telescope}{telescope for microlensing observations\xspace}
\newcommand{\tele}{telescope for microlensing observations on the ET satellite\xspace}

\usepackage[normalem]{ulem}
\usepackage[version=4]{mhchem}

\DeclareGraphicsExtensions{.pdf,.png,.jpg}

\title[]{Simulations of Triple Microlensing Events I: Detectability of a scaled Sun-Jupiter-Saturn System}

\input{author.tex}

\date{Accepted 2023 February 7. Received 2023 February 3; in original form 2022 November 23}
\pubyear{2022}

\begin{document}
\begin{CJK*}{UTF8}{gbsn}

\label{firstpage}
\pagerange{\pageref{firstpage}--\pageref{lastpage}}
\maketitle

\input{abstract}

\section{Introduction}\label{intro}

\input{intro_v2}

\section{Simulation and Light-curve Analysis}\label{sec:simu}

\input{simu}


\section{Results}\label{sec:result}
\input{result}

\section{Discussions}\label{sec:dis}
\input{dis}

\section{Conclusions}\label{sec:conclu}
\input{conclu}



\section*{Acknowledgements}
\addcontentsline{toc}{section}{Acknowledgements}
\textcolor{cRsp1}{We thank the anonymous referee for a valuable report that improved the paper.} R.K., W.Z., S.M., and J.Z. acknowledge support by the National Science Foundation of China (Grant No. 12133005). The authors acknowledge the Tsinghua Astrophysics High-Performance Computing platform at Tsinghua University for providing computational and data storage resources that have contributed to the research results reported within this paper. 


\section*{Data Availability}
The data underlying this article will be shared on reasonable request to the corresponding author.



\bibliographystyle{mnras}
\bibliography{main.bib}

\end{CJK*}
\bsp	
\label{lastpage}
\end{document}

%% file: author.tex
\author[R. Kuang et al.]{\parbox{\textwidth}{Renkun Kuang (匡仁昆) \orcid{0000-0003-2337-0533}$^{1,2}$,
\thanks{E-mail: krk18@mails.tsinghua.edu.cn}
Weicheng Zang (臧伟呈) \orcid{0000-0001-6000-3463}$^{1,3}$,
Shude Mao (毛淑德)$^{1,4}$,
Jiyuan Zhang (张纪元)$^{1}$,
Haochang Jiang (蒋昊昌) \orcid{0000-0003-2948-5614}$^{1,5}$}\vspace{0.4cm}\
\\
\parbox{\textwidth}{$^{1}$Department of Astronomy, Tsinghua University, Beijing 100084, China\\
$^{2}$Department of Engineering Physics, Tsinghua University, Beijing 100084, China\\
$^{3}$Center for Astrophysics $|$ Harvard \& Smithsonian, 60 Garden St.,Cambridge, MA 02138, USA\\
$^{4}$National Astronomical Observatories, Chinese Academy of Sciences, Beijing 100101, China\\
$^{5}$European Southern Observatory, Karl-Schwarzschild-Str 2, 85748 Garching, Germany\\
}}

%% file: abstract.tex
\begin{abstract}
Up to date, only 13 firmly established triple microlensing events have been discovered, so the occurrence rates of microlensing two-planet systems and planets in binary systems are still uncertain. With the upcoming space-based microlensing surveys, hundreds of triple microlensing events will be detected. To provide clues for future observations and statistical analyses, we initiate a project to investigate the detectability of triple-lens systems with different configurations and observational setups. As the first step, in this work we develop the simulation software and investigate the detectability of a scaled Sun-Jupiter-Saturn system with the recently proposed \textcolor{cRsp1}{\telescope on} the ``Earth 2.0 (ET)'' \textcolor{cRsp1}{satellite}. \textcolor{cRsp1}{With the same $\Delta\chi^2$ thresholds of detecting a single planet and two planets}, we find that the detectability of the scaled Sun-Jupiter-Saturn analog is about 1\% and the presence of the Jovian planet suppresses the detectability of the Saturn-like planet by $\sim $13\% regardless of the adopted detection $\Delta\chi^2$ threshold. This suppression probability could be at the same level as the Poisson noise of future space-based statistical samples of triple-lenses, so it is inappropriate to treat each planet separately during detection efficiency calculations.
\end{abstract}

%% file: intro_v2.tex
With \textcolor{mycolor1}{complementary} exoplanet detection methods such as the transit, radial velocity, microlensing, and direct imaging, there are more than 5,000 confirmed exoplanets\footnote{https://exoplanetarchive.ipac.caltech.edu/}. The fast pace of exoplanet detection triggers studies in various aspects from statistics (to capture the main characteristics of the known planetary sample) to planet formation and evolution theories (to understand and reproduce the main properties of the known planetary sample), see e.g., \cite{2021ZhuARAA} for a recent review. With these studies, we have a better understanding of both the solar system and exoplanet systems. For example, multi-body systems are common. Both \textcolor{mycolor1}{the} planet multiplicit\textcolor{mycolor1}{y} rate \citep[e.g.,][]{2022ZhuMultiplicity} and \textcolor{mycolor1}{the} stellar multiplicity rate of planet-host stars \citep[e.g.,][]{Wang2014ApJ} are substantial. These systems are valuable for studies on planet formation and evolution theories.

Among different exoplanet detection methods, the microlensing method \citep{Mao1991, Andy1992} allows us to detect exoplanets beyond the \textcolor{mycolor0}{snowline} and covers an important parameter space \citep[e.g.,][]{Gaudi2012, Mao2012}. Although some statistical studies \citep{mufun,Cassan2012,Wise,Suzuki2016} on microlensing planets showed that microlensing planets are abundant, with a frequency of order $\sim 50\%$--100\% (with planetary masses down to several Earth-mass), no specific study has been conducted to investigate the statistical properties of microlensing planets in multi-body systems. One reason is that the sample size of triple-lens systems still \textcolor{mycolor1}{being} small. There are 13 firmly established triple microlensing events, including five two-planet systems and eight \textcolor{mycolor1}{circumbinary (or circumstellar) planets} (see Table \ref{tab:alltriple} for the list). A larger sample can be built with the upcoming space missions such as the Nancy Grace Roman Space Telescope \citep[Roman, former\textcolor{mycolor1}{ly} WFIRST, ][]{Spergel2015}, the Chinese Space Station Telescope \citep[CSST, ][]{2022YanCSST}, and the ``Earth 2.0'' \citep[ET, ][]{Gould2021RAA, ET2} satellite\footnote{\textcolor{cRsp1}{The ET satellite consists of seven telescopes (with pupil diameter $\sim$30 cm), of which six will be used for transit observations and one for microlensing
observations.}}. For example, the joint survey \textcolor{cRsp1}{with the \tele} and the Korea Microlensing Telescope Network (KMTNet, \citealt{KMT2016}) is expected to discover several hundred planetary events, among them $\sim 5\%$\footnote{From the simulation results of \cite{Zhu2014ApJ}, as well as from the \textcolor{mycolor0}{apparent} multi-planetary fraction among all microlensing planets from observations.} \textcolor{mycolor1}{may} show multiple-planet signatures \citep{ET2}. 







Apart from the forecast that we will discover a large number of triple-lens systems, currently we know little about the properties of these systems. Many factors are involved such as the intrinsic property of the lenses and sources, the instruments, the observing strategies, and the detectabilities of triple-lens systems. To provide some clues for future observations, we initiate a project based on simulations to investigate the detectabilities of different triple-lens systems with different configurations \textcolor{mycolor0}{and} observation setups. \textcolor{cRsp1}{In addition, we note that} the magnification pattern and light curves from multiple planets \textcolor{mycolor0}{may be} degenerate with those from single-planet lensing \citep[e.g.,][]{Gaudi1998ApJ, 2014ZhuApJ2}\textcolor{mycolor0}{. Some of} the previous microlensing planet-detection sensitivity calculations ignore multiple planets or treat each planet as \textcolor{mycolor1}{being} independent \citep[e.g.,][]{OB141722}. \textcolor{mycolor0}{Here}, we investigate to what extent the detectability of a planet will be \textcolor{mycolor1}{suppressed or enhanced} by the presence of a second planet. Due to \textcolor{mycolor1}{the} large parameter space, in this paper we focus on building \textcolor{mycolor0}{a} pipeline from light curve calculation to planetary signal detection. \textcolor{mycolor1}{And as the first step, we} take a scaled Sun-Jupiter-Saturn system as the lens and take the \textcolor{cRsp1}{\tele} \textcolor{mycolor0}{as an example survey. Below we discuss why} we choose a scaled Sun-Jupiter-Saturn system as the lens. 


Despite the discovery of many exoplanets, an outstanding question remains to be answered from both theoretical and observational sides is how unique the solar system is \citep[e.g.,][]{2004Beer, 2015Martin, 2019Portegies, 2020Raymond}, i.e., what is the frequency of solar system analogs?

The formation and evolution of the solar system involve\textcolor{mycolor1}{s} many intertwined processes, especially when multiple planets cover a wide range of orbital distances. The \textcolor{mycolor1}{in-situ} formation of outer giant planets is difficult due to their longer formation timescale compared with the lifetime of protoplanetary disks. \textcolor{mycolor1}{In addition}, the solar system has some peculiarities, such as the Earth/Mars mass ratio being exceptionally large, and the asteroid and Kuiper belts being low mass yet dynamically excited \citep{2020Raymond}. Some studies found that the instability of giant planets may have a significant impact on the inner solar system \citep[e.g.,][]{2019Clement}. The Grand Tack model \citep{2011Walsh} suggested that Jupiter and Saturn underwent a two-stage, inward-then-outward migration, which can truncate the inner disk of rocky material and can explain the large Earth/Mars mass ratio.

Due to the important role of Jupiter and Saturn in the formation of the solar system, and the dimensions of a planetary system may scale with the snowline radius \citep{Min2011, snowline}, throughout this paper we define the solar system analogs as \textcolor{mycolor0}{systems that have} both Jovian and Saturn-like planets at similar locations in units of the snowline radius of the host star. Under this definition, the determination of the frequency of solar system analogs from the observational side is still challenging due to the difficulties in detecting long-period planets. For example, a decade-long radial velocity survey would be required to detect Jupiter orbiting the Sun, while Saturn, Uranus, and Neptune are too distant \citep{2020Raymond}. 


Among the five firmly established two-planet systems detected with microlensing, OGLE-2006-BLG-109L has a pair of Jupiter/Saturn planets and was regarded as a ``scaled version'' of our solar system \citep{OB06109, OB06109_Dave}. It is not clear whether this detection \textcolor{mycolor1}{is} due to a coincidence or due to the intrinsic \textcolor{mycolor1}{high} fraction of such systems. This leads to one of the aims of this paper: \textcolor{mycolor1}{to investigate} the detectability of \textcolor{mycolor1}{scaled Sun-Jupiter-Saturn systems} with \textcolor{mycolor1}{a space} telescope, \textcolor{mycolor1}{such as ET}.

There are some other reasons for us to use a scaled Sun-Jupiter-Saturn system as the lens. First, in most cases \textcolor{mycolor1}{only} two planets are detectable although there are multiple planets in the lenses \citep{Shvartzvald_Maoz2012, Zhu2014ApJ}. Second, the inner and the outer planetary systems are correlated that outer cold Jupiters almost always have inner planetary companions \citep{2021ZhuARAA}, so the frequency of Jupiter/Saturn analogs could be regarded as a rough approximation to the frequency of solar system analogs. Third, the multiplicity rate of giant planets is about 50\% \citep{2016Bryan, 2022ZhuMultiplicity}. Last\textcolor{mycolor1}{ly}, the light curve calculations for multiple lenses are computationally expensive.



Ever since the microlensing planet detection method has been proposed, \textcolor{mycolor1}{parallel} with observational efforts, there were numerical simulations to investigate the detection probabilities and to predict possible outcomes of observational experiments. Due to it being computationally expensive in both light curve (or magnification map) calculations and analys\textcolor{mycolor1}{es (to} identify planetary / multiplanetary events), previous simulations generally investigate \textcolor{mycolor1}{either} with small samples of lenses \citep[e.g.,][]{Gaudi1998ApJ} or source trajectories \citep[e.g.,][]{Zhu2014ApJ, 2014ZhuApJ2}, or with simplified ``detection criteria'' \citep[e.g.,][]{Ryu2011}. \textcolor{mycolor0}{We survey these works below.}

\cite{Ryu2011} investigated the detection probability of a low-mass planet from high-magnification events caused by triple-lens systems composed of a star, a Jovian mass planet, and a low-mass planet. They use the \cite{Andy1992} criterion \textcolor{mycolor0}{to calculate} the fractional deviation in magnification maps at the central region. \textcolor{mycolor1}{However,} the detection of \textcolor{mycolor0}{deviation} signals for a low-mass planet may not be sufficient for the discovery of such a planet.

\cite{Shvartzvald_Maoz2012} used scaled solar system analogs as the lenses and took sampling sequences and photometric error distributions from a real observing experiment. They found that a generation-\uppercase\expandafter{\romannumeral2} microlensing network can find about 50 planetary events, among them $1/6$ reveal two-planet anomalies. \textcolor{mycolor1}{However,} they did not fit \textcolor{mycolor1}{their} light curve\textcolor{mycolor1}{s} with \textcolor{mycolor1}{a} binary-lens model to identify the two-planet events. Instead, they use a ``running'' local $\chi^2$ estimator with 31 consecutive points to detect \textcolor{cRsp1}{short-timescale} deviations in the light curve relative to a point-lens model (see their \S3.5 and \S4.3 for more details). \textcolor{mycolor1}{Furthermore}, they did not investigate how the detectability of a planet will be \textcolor{mycolor1}{inhibited or promoted} by other planets in the same system.

\cite{Zhu2014ApJ, 2014ZhuApJ2} conducted a simulation in the context of the core accretion planet formation model \citep{Ida2004}. \textcolor{cRsp1}{\cite{Zhu2014ApJ} generated} 10 light curves for each of the 669 selected planetary systems and found the fraction of planetary events is 2.9\%, out of which 5.5\% show multiple-planet signatures. Among the 23 two-planet event candidates, they confirmed 16 triple-lens events with the criterion $\Delta\chi^2>300$, where $\Delta\chi^2$ is the $\chi^2$ difference between the best-fit binary-lens model and the input multiplanetary model. 



\textcolor{mycolor1}{To remedy the small sample and loose detection critetia of previous studies,} in this work, we simulate and analyse $4\times10^{5}$ events to investigate the detectability of a scaled Sun-Jupiter-Saturn
system, the suppression and enhancement effects, and their dependence on \textcolor{mycolor0}{parameters,} e.g., the impact parameter. The paper is structured as follows. In \S\ref{sec:simu}, we introduce the microlensing basics, and details about how we generate light curves and detect planets from the light curves. In \S\ref{sec:result}, we present the results. We then discuss the results and future works in \S\ref{sec:dis} \textcolor{mycolor1}{and conclude} in \S\ref{sec:conclu}. 

%% file: simu.tex
\subsection{\textcolor{mycolor1}{Microlensing basics}}

A microlensing event \textcolor{mycolor1}{occurs} when a lens object is close to the line of sight from the observer to the background source. The light rays from the background source are deflected by the lens, forming multiple images as seen by the observer. The projected position of the source, lens, and images are related by the lens equation \citep{Witt1990},
\begin{equation}
z_s = z - \sum_{j=1}^{N}\frac{m_j}{\overline{z}-\overline{z_j}},
\label{equ:lens_equ}
\end{equation}
where $z_s, z, z_j$ are respectively the complex positions of the source, the image, and the $j$-th lens, $\overline{z}$ and $\overline{z_j}$ are respectively the complex conjugates of $z$ and $z_j$, and $m_j$ is the fractional mass of the $j$-th lens, with $\sum_jm_j=1$. The positions are represented in \textcolor{mycolor0}{units of the} angular Einstein radius $\theta_{\rm E}$ of the lensing system,
\begin{equation}
\begin{aligned}
\theta_{\rm E} &= \sqrt{\frac{4GM}{c^2}\frac{D_{\rm S} - D_{\rm L}}{D_{\rm L}D_{\rm S}}}\\
& = 0.55\; \mathrm{mas} \sqrt{\frac{1-D_{\rm L}/D_{\rm S}}{D_{\rm L}/D_{\rm S}}} \left( \frac{8\; \mathrm{kpc}}{D_{\rm S}} \frac{M}{0.3\; M_{\odot}} \right)^{1/2},
\end{aligned}
\label{equ:thetaE}
\end{equation}
where $M$ is the total mass of the lens system, $D_{\rm L}$ and $D_{\rm S}$ are the distances from the observer to the lens and the source, respectively, $c$ is the speed of light, and $G$ is the gravitational constant.

In Equation (\ref{equ:lens_equ}), for a given source position $z_s$, multiple solutions for $z$ exist. For an extended source, multiple distorted images are formed, with separations of the order of $\mathrm{mas}$. In most cases, these images cannot be resolved. Instead, the observable is the \textcolor{mycolor1}{change of the} total flux \textcolor{mycolor1}{due to the changing magnification}. The time scale of a microlensing event \textcolor{mycolor1}{is} relate\textcolor{mycolor1}{d} to the lens-source relative proper motion $\mu_{\rm rel}$ by
\begin{equation}
t_{\rm E} = \frac{\theta_{\rm E}}{\mu_{\rm{rel}}}.
\end{equation}

Throughout the paper, we denote a triple-lens system with five parameters ($s_2$, $q_2$, $s_3$, $q_3$, $\psi$). Here $q_2$ and $q_3$ are respectively the mass ratios of the second and third lens to the first lens, $s_2$ and $s_3$ are respectively the separations normalized to $\theta_{\rm E}$ of the second and third lens from the first lens, and $\psi$ represents the orientation angle of the third lens. The relations between ($s_2$, $q_2$, $s_3$, $q_3$, $\psi$) and $m_j$, $z_j$ are,
\begin{equation}
\begin{aligned}
&m_1 = 1/(1+q_2+q_3),\quad m_2 = q_2m_1, \quad m_3 = q_3m_1,\\
&z_1 = -q_2s_2/(1+q_2)+i\;0,\\
&z_2 = s_2/(1+q_2)+i\;0,\\
&z_3 = -q_2s_2/(1+q_2)+s_3\cos{\psi} + i\;s_3\sin{\psi},\\
\end{aligned}
\label{equ:coord}
\end{equation}
\textcolor{mycolor0}{where $i$ is the imaginary unit. Note that the first and second lenses are on the $x$-axis.}

\subsection{Lens and source properties}

In reality, the lens physical parameters vary from event to event. For simplicity, we fix the lens mass as $M=0.5\, M_{\odot}$, the lens distance as $D_{\rm L} = 5\; \mathrm{kpc}$, the source distance as $D_{\rm S} = 8\; \mathrm{kpc}$. So the angular Einstein radius is $\theta_{\rm E} = 0.55$ $\mathrm{mas}$, corresponding to an Einstein radius of $R_{\rm E} = D_{\rm L}\theta_{\rm E} = 2.75$ au. 

In the Solar system, the mass ratios of Jupiter and Saturn to the Sun are $q_{\rm Jupiter} = 9.546\times 10^{-4}$, and $q_{\rm Saturn} = 2.857 \times 10^{-4}$, respectively. The distances from Jupiter and Saturn to the Sun are $4.965\; \mathrm{au}$ and $9.905\; \mathrm{au}$, respectively\footnote{The distances are calculated based on real-time locations at the epoch $\rm{MJD}=59731$ with the Python package \textcolor{cRsp1}{\href{https://github.com/IoannisNasios/solarsystem}{\texttt{solarsystem}} \citep{Nasios2020}}.}. We adopt $q_2 = q_{\rm Jupiter}$ and $q_3 = q_{\rm Saturn}$. The planet-host distances are scaled such that the distances remain the same in units of the \textcolor{mycolor0}{snowline} distance\footnote{We assume the \textcolor{mycolor0}{snowline} distance scales as $a_{\rm SL} = 2.7(M/M_{\odot})$~{\rm au}, \citep{snowline}.}, so (regardless of the projection effect)
\begin{equation}
s_{2,\rm{true}} = 0.903, \quad s_{3,\rm{true}} = 1.801. 
\end{equation}

\begin{figure}
    \centering
    \includegraphics[width=\columnwidth]{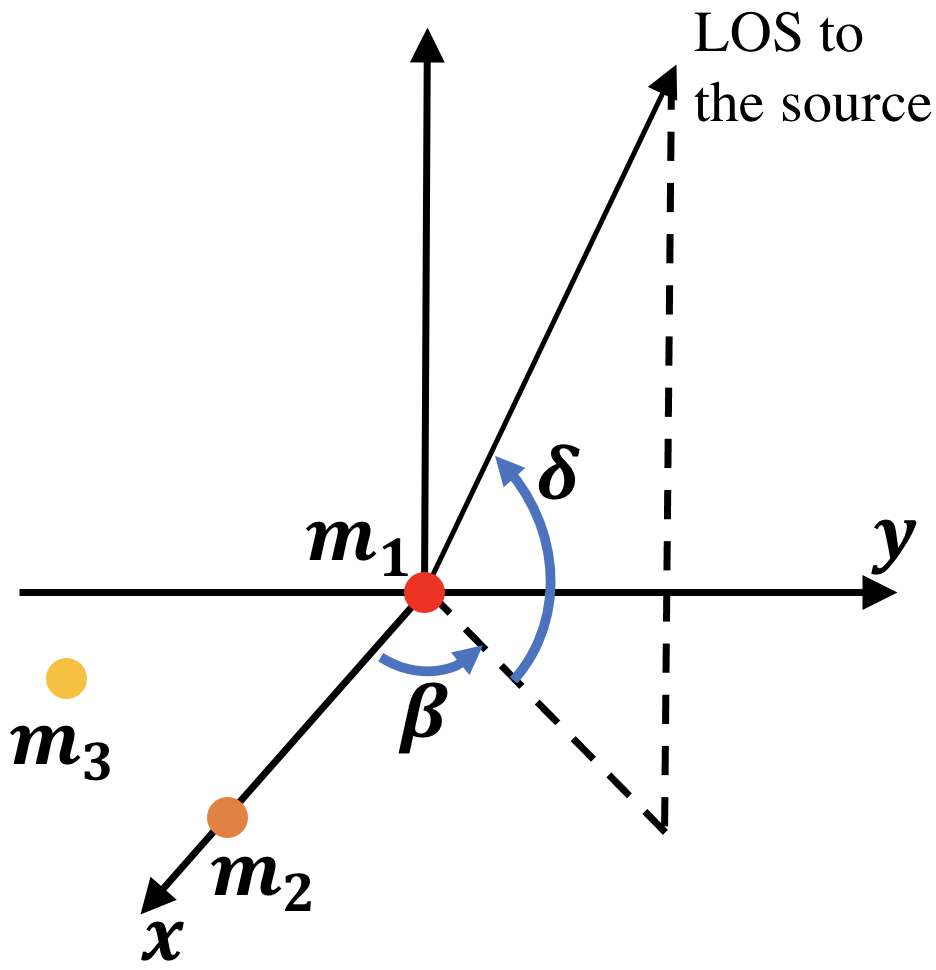}
    \caption{A Schematic diagram shows the geometry of the lens orbital plane and the direction of the source. The host star $m_1$ and the two planets $m_2$ and $m_3$ are located \textcolor{mycolor0}{in} the same plane\textcolor{mycolor0}{, where $m_1$ and $m_2$ are on the $x$-axis}. The two angles $\beta$ and $\delta$ define the direction of the line-of-sight (LOS) to the source.}
    \label{fig:geom}
\end{figure}

We randomly generate 200 triple-lens systems by using the above lens parameters but with different planet orbital phases and \textcolor{mycolor0}{viewing angles}. We place the planets on their orbit\footnote{Jupiter and Saturn are close to the 5 : 2 resonance (\textcolor{mycolor0}{so-}called mean-motion near resonance). They obey the relation $5n_{\rm{Saturn}} - 2n_{\rm{Jupiter}} \approx 0$, where $n$ is the inverse of orbital period. But an exact resonance does not exist for Jupiter and Saturn \citep{Michtchenko2001}. So we do not consider the mean motion resonance and \textcolor{cRsp1}{treat the two planets as having} random orbital phases.} by sampling the position angle $\psi_{\rm{true}}$ \textcolor{mycolor1}{from} a uniform distribution $\mathcal{U}(0,2\pi)$\textcolor{mycolor1}{, and the} orbital plane of the triple-lens system at random directions by using two directional angles, $\beta$ and $\delta$. They define the direction of the source as seen from the orbital plane of the lens system. Figure \ref{fig:geom} shows the geometry. We randomly choose $\beta$ from a uniform distribution $\mathcal{U}(0,2\pi)$, and $\sin\delta$ from a uniform distribution $\mathcal{U}(-1,1)$. For each \textcolor{mycolor1}{combination} of ($\psi_{\rm{true}}$, $\beta$, $\delta$), we can calculate the projected positions of $m_2$ and $m_3$, and convert their positions to triple-lens parameters ($s_2$, $s_3$, $\psi$) which follow our convention defined in Equation (\ref{equ:coord}).

For a given triple-lens system, we then generate 2000 source trajectories with random impact parameter $u_0$ and direction angle $\alpha$, so there are $200\times2000 = 4\times 10^5$ lens-source pairs, with the same $t_0$ and $t_{\rm{E}}$, where $t_0$ is the time of closest approach of the source to the origin, $u_0$ is the closest distance of the source to the origin in units of $\theta_{\rm E}$. We use $t_0=9000$ (in units of $\rm{HJD}^{\prime} \equiv \rm{HJD} - 2450000$) and $t_{\rm E} = 30$ days, corresponding to a lens-source relative proper motion $\mu_{\rm{rel}} = 6.7\, \rm{mas}/\rm{yr}$. We sample the trajectory angle $\alpha$ from a uniform distribution $\mathcal{U}(0,2\pi)$. \textcolor{cRsp1}{Because events with small $u_0$ (i.e., high-magnification events) are more sensitive to planets \citep{Griest1998}, we use the importance sampling method by drawing $\log u_0$ from a uniform distribution of $\mathcal{U}(-3,0.3)$, rather than directly drawing $u_0$ from a uniform distribution of $\mathcal{U}(10^{-3},2)$. Then, for $M$ detectable events from $N(=2000)$ simulated light curves, the detectability is 
\begin{equation}
P = \frac{\Sigma_{j=1}^{M} u_{0,j}}{\Sigma_{i=1}^{N} u_{0,i}},
\label{equ:equprob}
\end{equation}
where $i=1,\cdots,N$ represents the index for all source trajectories and $j=1,\cdots,M$ represents the index for all source trajectories that allow the detection of two planets.}


We adopt a source star with a radius $R_{*}=0.95\, R_{\odot}$, i.e., the scaled source size $\rho = \theta_{*}/\theta_{\rm E}=0.001$. We assume that the source has a uniform surface brightness profile, i.e., we do not consider the limb darkening effect. \textcolor{mycolor1}{Furthermore, for simplicity}, we do not account for the lens orbital motion \citep{MB09387, OB09020} and the microlens parallax effect \citep{Gould2000} in this work, \textcolor{cRsp1}{although these two effects have been detected in the two-planet event, OGLE-2006-BLG-109 \citep{OB06109, OB06109_Dave}}. 

\subsection{Generating light curves}\label{sec:lkvgen}

\begin{figure}
    \centering
    \includegraphics[width=\columnwidth]{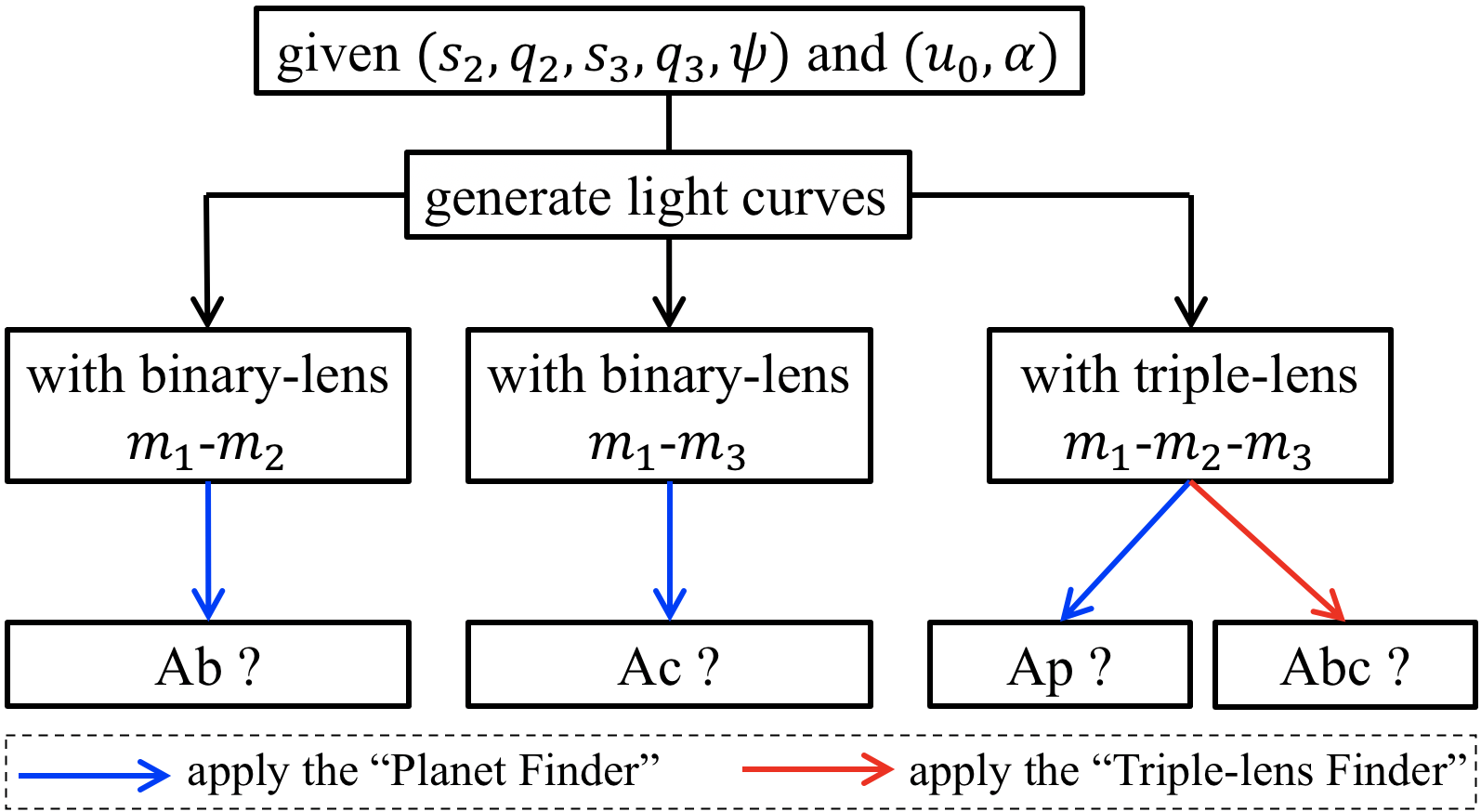}
    \caption{A flowchart shows the overall procedures used to achieve the main goals of this work. We follow the convention in the microlensing field to represent the lensing object of an event. The host star is represented with capital letter ``A'', followed by small letters represent\textcolor{mycolor0}{ing} the planets. See \S\ref{sec:result} for the detailed definitions of the symbols in this flowchart.}
    \label{fig:flowchart}
\end{figure}

\begin{figure*}
    \centering
    \includegraphics[width=\textwidth]{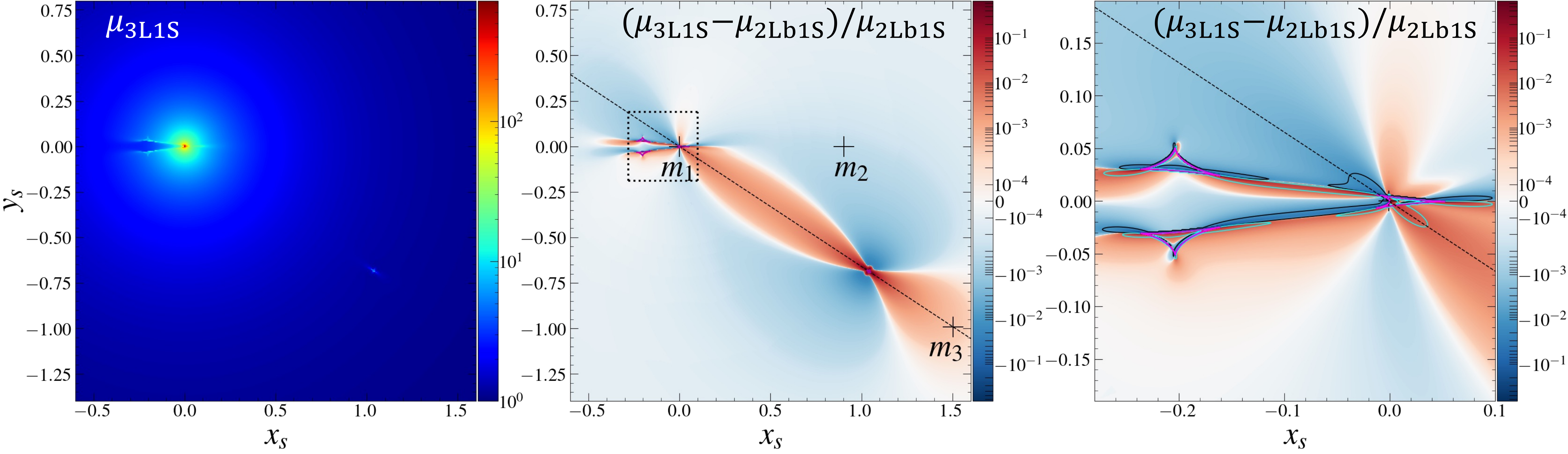}
    \caption{An example shows \textcolor{cRsp1}{the influence of $m_3$ on the magnification pattern generated by the binary-lens system $m_1$-$m_2$.} The left panel shows the magnification map corresponding to the triple-lens system ($\mu_{\rm{3L1S}}$). The middle and right panels show the fractional difference (magnification excess) $(\mu_{\rm{3L1S}} - \mu_{\rm{\textcolor{cRsp2}{ 2Lb1S}}})/\mu_{\rm{\textcolor{cRsp2}{2Lb1S}}}$, the dashed black line connects $m_1$ and $m_3$, the solid magenta curve shows the caustics of the triple-lens system. The crosses in the middle panel show the locations of the three lenses. In the right panel, the solid cyan and black lines show the $\pm1\%$ contours \textcolor{mycolor1}{for the dotted region showed in the middle panel}. The mass ratios and separations (in units of $\theta_{\rm{E}}$) of these two planets to the host star are $(s_2,q_2)=(0.903, 9.55\times10^{-4})$, $(s_3,q_3)=(1.80, 2.86\times10^{-4})$, respectively. The position angle of $m_3$ is $\psi = -33.4^{\circ}$.}
    
    \label{fig:egmagmap}
\end{figure*}

\begin{figure*}
    \centering
    \includegraphics[width=\textwidth]{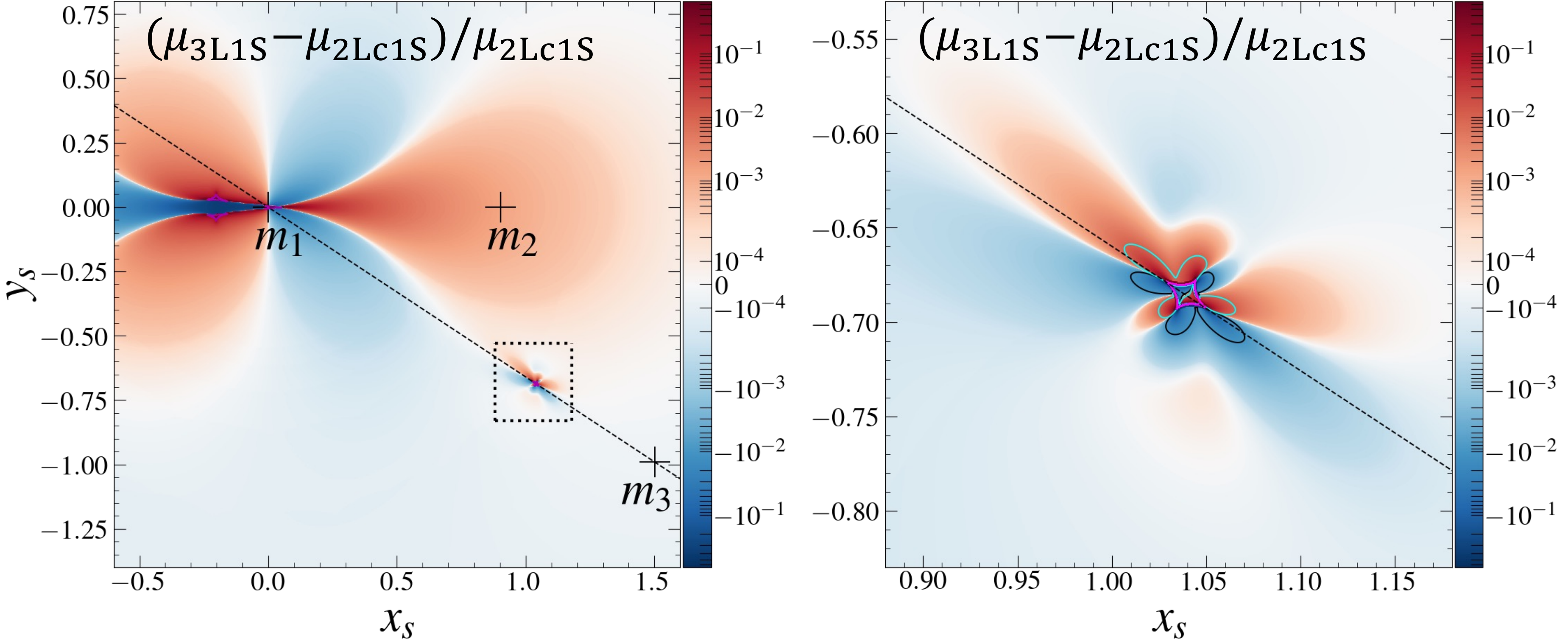}
    \caption{The same as Figure \ref{fig:egmagmap}, except that now we are showing the influence of $m_2$ on the magnification pattern generated by the binary-lens system $m_1$-$m_3$. The mass ratios and separations of these two planets are the same as in Figure \ref{fig:egmagmap}. \textcolor{mycolor1}{The right panel shows a magnified view of the dotted region in the left panel.}}
    
    \label{fig:egmagmap_2}
\end{figure*}

The first goal of this work is to investigate the detectability of Jupiter/Saturn analogs. 
The second goal is to find out how the two planets would affect each other's detectability. The influence of $m_3$ on the magnification pattern generated by $m_1$-$m_2$ is non-negligible, see Figure \ref{fig:egmagmap} for an example map of magnification excess \citep[\textcolor{mycolor1}{see also}][]{Chung2005}. The planet $m_3$ with mass ratio $q_3 = 2.857\times10^{-4}$ perturbs the magnification pattern of the $m_1$-$m_2$ system not only along the $m_1$-$m_3$ axis, but also along the $m_1$-$m_2$ axis (horizontal). \textcolor{mycolor1}{In turn}, the magnification pattern of the $m_1$-$m_3$ system is also influenced by the presence of $m_2$, as shown in Figure \ref{fig:egmagmap_2}. The overall procedures used to achieve the above two goals are shown as a flowchart in Figure \ref{fig:flowchart}. 


In the following, we introduce the detail of the light curve calculations within the context of the ET satellite. \textcolor{cRsp1}{The \tele will have} a 4 $\mathrm{deg}^2$ field of view, and will monitor the Galactic bulge with a cadence of 10 minutes from March 21 to September 21 every year. We take 26.8 as the $I$-band magnitude zero-point (1 count/s). For simplicity, we adopt the background magnitude due to \textcolor{mycolor1}{surrounding} stars as \textcolor{mycolor1}{$I_{\rm{blend}}$} = 18.9 and take 20.9 as the apparent $I$-band magnitude of the source star. We assume the noise is dominated by \textcolor{mycolor1}{Possion} noise, and the standard deviation of each photometric measurement is the square root of the flux measurement in photon counts. \textcolor{mycolor1}{To be conservative, the chosen star is quite faint and thus has fairly large noise.} We list the simulation-related parameters in Table \ref{tab:lens_source}. 




For a given set of lensing parameters $(s_2, q_2, s_3, q_3, \psi)$ we generate three light curves for each source trajectory $(u_0,\alpha)$. For the first light curve, the lensing object is a binary-lens system with lensing components $m_1$-$m_2$ (designated as the \textcolor{cRsp2}{2Lb1S} light curve). For the second light curve, the lensing components are $m_1$-$m_3$ (designated as the \textcolor{cRsp2}{2Lc1S} light curve). For the third light curve, the lensing object is the triple-lens system, $m_1$-$m_2$-$m_3$ \textcolor{cRsp1}{(designated as the 3L1S light curve)}. \textcolor{mycolor1}{N}ote that when calculating the three light curves, we take into account the fact that $\thetae$ changes as the total mass of the lens changes.


We use the \texttt{VBBinaryLensing} \citep{Bozza2010, Bozza2018} and \texttt{triplelens}\footnote{\textcolor{cRsp1}{We used version 1.0.8 with the Github commit ID: \href{https://github.com/rkkuang/triplelens/tree/bb348cfad1ea865ba5f533f5dbba71c78a66dc81}{bb348cfad1ea865ba5f533f5dbba71c78a66dc81}.}} \citep{2021Renkun} to calculate the magnifications for binary-lens and triple-lens, respectively. Then the light curves are calculated by multiplying the magnifications with the baseline source flux and adding the constant blend flux and measurement noises. So for each light curve, we can calculate the $\chi^2_{\rm{theoretical}}$ by using the theoretical model.

\begin{table}
    \renewcommand\arraystretch{1.25}
    \caption{\textcolor{mycolor1}{The simulation-related parameters}.}
    
    \begin{tabular}{cc|cc}
    \hline
    Parameters & Value & Parameters & Value \\
    \hline
    $q_2 (10^{-4})$ & $9.546$ & $\psi_{\rm{true}}$ & $\mathcal{U}(0,2\pi)$\\
    $q_3 (10^{-4})$ & $2.857$ & $\beta$ & $\mathcal{U}(0,2\pi)$\\
    $s_{2,\rm{true}}$ & $0.903$ & $\sin\delta$ & $\mathcal{U}(-1,1)$ \\
    $s_{3,\rm{true}}$ & $1.801$ & $\log u_0$ & $\mathcal{U}(-3,0.3)$\\ 
    \textcolor{mycolor1}{$I_{\rm{blend}}$} & $18.9$ & $\te$ (days) & $30.0$\\
    $I_{\rm{source}}$ & $20.9$ & $\rho (10^{-3})$ & $1.0$ \\    
    $I_{\rm{zero point}}$ & $26.8$ & $\alpha$ & $\mathcal{U}(0,2\pi)$ \\

    
    
    \hline
    \end{tabular}
    \label{tab:lens_source}
\end{table}

\begin{figure*}
    \centering
    \includegraphics[width=\textwidth]{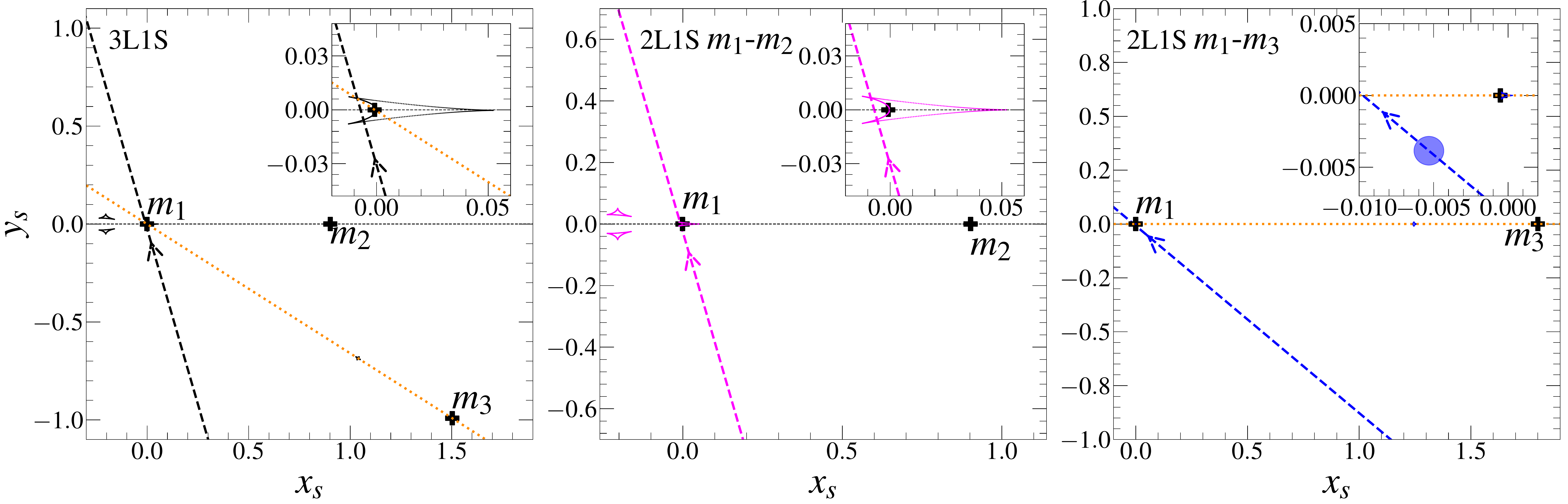}
    \caption{An example case of caustic structures and source trajectories. From left to right, the \textcolor{cRsp1}{black, magenta, and blue solid} curves are the caustics corresponding to the lens $m_1$-$m_2$-$m_3$, $m_1$-$m_2$, and $m_1$-$m_3$, respectively.  The \textcolor{cRsp1}{dotted black line shows the $m_1$-$m_2$ axis, and the} dotted orange line shows the $m_1$-$m_3$ axis. \textcolor{cRsp1}{The black plus signs represent the lensing objects. In the inset of the right panel, the filled blue circle represents the source.} The black, magenta, and blue \textcolor{cRsp1}{dashed} lines with arrows show the source trajectories corresponding to each lens system. The corresponding model light curves are shown in the upper panel of Figure \ref{fig:eglkv} with the same colour.}
    \label{fig:eglkv_caus}
\end{figure*}

\begin{figure}
    \centering
    \includegraphics[width=\columnwidth]{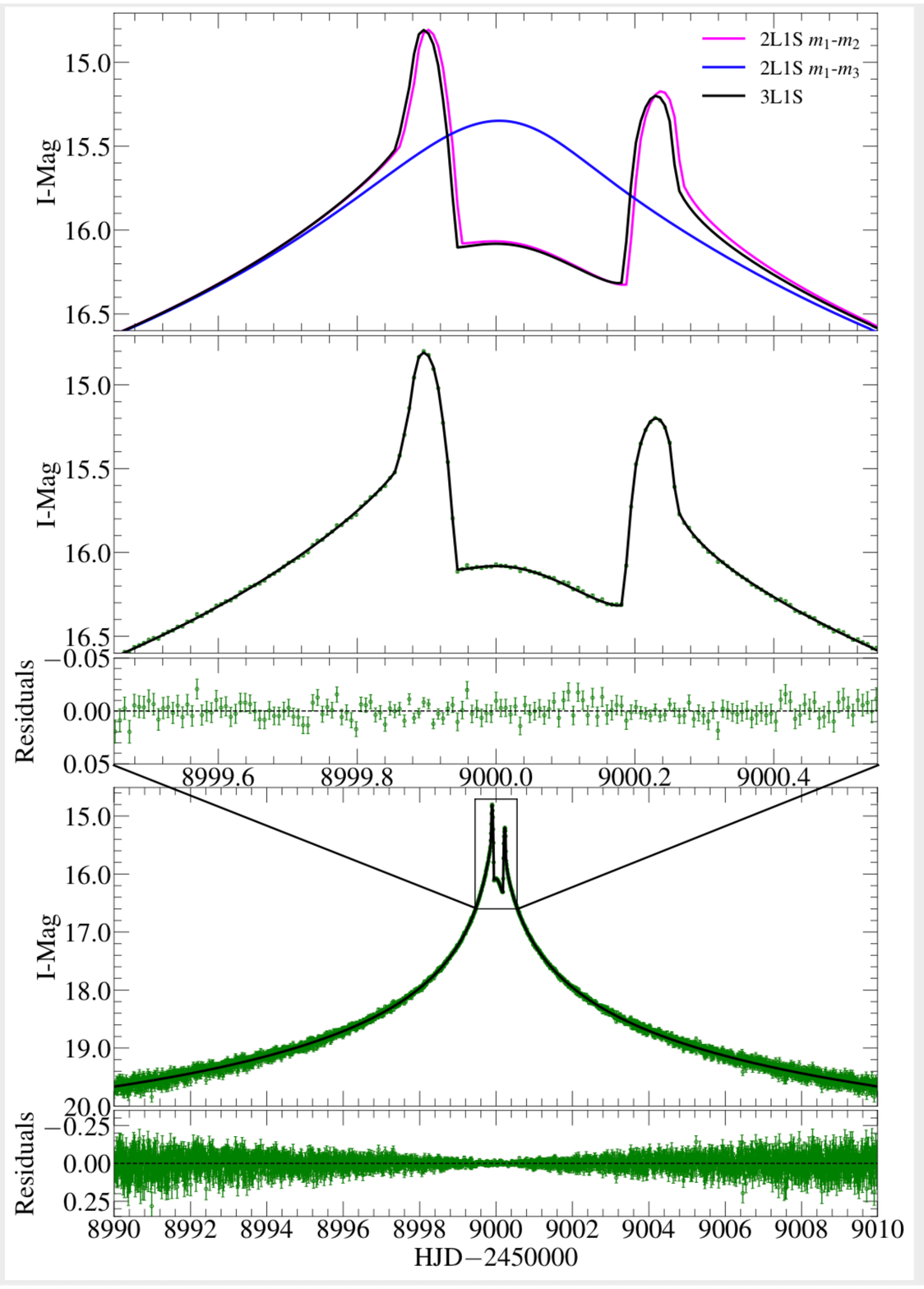}
    \caption{The corresponding light curves for the example case\textcolor{mycolor0}{s as} shown in Figure \ref{fig:eglkv_caus}. The upper panel shows the three model light curves with different colours. The \textcolor{cRsp1}{green} points with error bars in the middle and bottom panels are the simulated data points corresponding to the triple-lens.}
    \label{fig:eglkv}
\end{figure}

Figure \ref{fig:eglkv_caus} shows an example of caustic structures and source trajectories. From left to right, the red curves are the caustics corresponding to the lens $m_1$-$m_2$-$m_3$, $m_1$-$m_2$, and $m_1$-$m_3$, respectively. The dotted orange line shows the $m_1$-$m_3$ axis. The black, magenta, and blue \textcolor{cRsp2}{dashed} lines with arrows show the source trajectories corresponding to each lens system. The corresponding model light curves are shown in the upper panel of Figure \ref{fig:eglkv} with the same colours. The \textcolor{cRsp2}{green} points with error bars in the middle and bottom panels of Figure \ref{fig:eglkv} are the simulated data points corresponding to the triple-lens.

\subsection{Planet Finder}

The aim of this step is to determine from each of the three light curves, whether there are detectable planetary signatures. 

We fit each of the three light curves with the single-lens-single-source (1L1S) model. The free (non-linear) parameters are $t_0$, $u_0$, $t_{\rm E}$, and $\rho$, their initial values are taken from the values used to generate the corresponding light curves. We use the Nelder-Mead simplex algorithm \citep{NeldMead65, NelderMead2012} to find the best-fitting parameters. 

\textcolor{cRsp1}{For clarity, we use the following notation to represent the $\chi^2$ differences between different models. For $\Delta \chi^2_{\rm A, B} = \chi^2_{\rm A} - \chi^2_{\rm B}$, we use model B to generate data points in the light curve and calculate the theoretical $\chi^2$ $(\chi^2_{\rm B})$, and fit the data points with model A to obtain $\chi^2_{\rm A}$.}

The criterion we used to evaluate whether a light curve contains planetary signal is that $\Delta \chi_{\rm{1L1S\textcolor{cRsp1}{,theoretical}}}^2 = \chi_{\rm{1L1S}}^2 - \chi^2_{\rm{theoretical}} > 200$\textcolor{cRsp1}{, here ``theoretical'' can be 2L1S or 3L1S models. We note that the $\Delta \chi_{\rm{1L1S,2L1S}}^2$ thresholds vary \textcolor{cRsp2}{between} different statistical studies, simulations, and planetary searches. The current four microlensing planetary statistical studies used $\Delta \chi_{\rm{1L1S,2L1S}}^2$ thresholds from 100 to 500 \citep{mufun,Cassan2012,Wise,Suzuki2016}. The simulation conducted by \cite{Zhu2014ApJ} and \cite{Henderson2014} adopted $\Delta \chi_{\rm{1L1S,2L1S}}^2 = 200$ or 500, respectively. The lowest $\Delta \chi_{\rm{1L1S,2L1S}}^2$ of the systematic KMTNet planetary anomaly search is 60 \citep{OB191053,2019_prime}. We adopt 200 as} a preliminary threshold to investigate the related probabilities and will explore the dependence of the results on the $\Delta\chi^2$ threshold in \S\ref{sec:dep_on_dchisq}.

\subsection{Triple-lens Finder}\label{sec:triple_finder}
This step aims to evaluate whether $m_2$ and $m_3$ can be simultaneously detected from a 3L1S light curve, i.e., whether a 2L1S model is sufficient to explain the 3L1S light curve.\footnote{We do not consider the scenario that other models \citep[e.g., binary-lens binary-source,][]{OB141722,OB180532,KB191953,KB210240} could be degenerate with triple-lens model.} If 1L1S and 2L1S models \textcolor{mycolor1}{cannot} explain the light curve, we regard this as the case that both planets are detectable. We fit the 3L1S light curve with a 2L1S model, the initial guess for the 2L1S model parameters is determined by the following steps. 1). For the 2L1S model consists \textcolor{mycolor1}{of} $m_1$-$m_2$, \textcolor{mycolor1}{we} calculate the initial chi-squared $\chi^2_{\rm{init, m_2}}$. 2). For the 2L1S model consists \textcolor{mycolor1}{of} $m_1$-$m_3$, \textcolor{mycolor1}{we} calculate the initial chi-squared $\chi^2_{\rm{init, m_3}}$. 3). We choose the 2L1S model with \textcolor{mycolor1}{the} smaller $\chi^2$ as the initial 2L1S parameters for \textcolor{mycolor1}{further minimisation with} the Nelder-Mead algorithm. 

The criterion we used to evaluate whether a light curve contains two-planet signals is $\Delta \chi_{\rm{1L1S\textcolor{cRsp1}{,3L1S}}}^2 > 200$ \textcolor{mycolor1}{and} $\Delta \chi_{\rm{2L1S\textcolor{cRsp1}{,3L1S}}}^2 > 200$, i.e., the planetary signal can be detected with the ``Planet Finder'', in the meanwhile the two-planet signal can be detected with the ``Triple-lens Finder'' \textcolor{cRsp1}{from the triple-lens light curve. We note that there are fewer 2L1S events than 1L1S events. For example, the 2L1S rate of the systematic KMTNet planetary anomaly search is about 11\% \citep{2019_prime,2018_prime,2018_subprime}. Thus, statistically the $\Delta \chi_{\rm{2L1S,3L1S}}^2$ threshold should be lower than the $\Delta \chi_{\rm{1L1S,2L1S}}^2$
threshold. However, in practice as pointed out by \cite{Zhu2014ApJ},  with a much lower occurrence of 3L1S than that of 2L1S, the deviations after subtraction of the first planet may be explained by systematics in the data (e.g., \citealt{OB130341}), stellar variabilities or high-order effects (the lens orbital motion and microlens parallax effect). Thus, we may need a higher $\Delta \chi_{\rm{2L1S,3L1S}}^2$ threshold. Because this question is still uncertain, in this work we adopt $\Delta \chi_{\rm{1L1S,2L1S}}^2 = \Delta \chi_{\rm{2L1S,3L1S}}^2$.}


%% file: result.tex


For a given set of $(s_2, q_2, s_3, q_3, \psi)$, different source trajectory parameter $(u_0, \alpha)$ determines the shape of the corresponding light curves \textcolor{cRsp1}{and determines whether planet $m_2$ and/or $m_3$ can be detected.} Some source trajectories allow the detection of both planets in the triple-lens light curve, we denote these events as ``Abc''. \textcolor{cRsp1}{The flowchart in Figure \ref{fig:flowchart} is an overview on how ``Abc'' events and other type of events (``Ab'', ``Ac'' and ``Ap'') are found with ``Planet Finder'' and/or ``Triple-lens Finder''. We explain them in more detail as follows,} 
\begin{itemize}
\item ``Ab'': The Jovian planet is detectable in the \textcolor{cRsp2}{2Lb1S} light curve (generated with the lensing components $m_1$-$m_2$, see \S\ref{sec:lkvgen} for details) The ``Planet Finder'' is used to find these events by fitting the 1L1S model to the \textcolor{cRsp2}{2Lb1S} light curve to check whether $\Delta\chi_{\rm 1L1S, \textcolor{cRsp2}{2Lb1S}}^2>200$.
\item ``Ac'': The Saturn-like planet is detectable in the \textcolor{cRsp2}{2Lc1S} light curve. The ``Planet Finder'' is used to find these events. 
\item ``Ab $\cap$ Ac'': Both planets are detectable from the corresponding 2L1S light curves. \textcolor{cRsp1}{These events are selected as the intersection of the above two samples. These were} considered two-planet event candidates in \cite{2014ZhuApJ2}, \textcolor{mycolor1}{however,} planets can be individually detected does not guarantee that they can \textcolor{mycolor0}{be simultaneously} detected in the same event.
\item ``Ap'': The planet signal is detectable in the 3L1S light curve. \textcolor{cRsp1}{The ``Planet Finder'' is used to find these events by fitting the 1L1S model to the 3L1S light curve to check whether $\Delta\chi_{\rm 1L1S, 3L1S}^2>200$.} If the detectable planet is $m_2$, we denote the event as ``$\rm{Ap,b}$''. \textcolor{cRsp1}{Both ``Planet Finder'' and ``Triple-lens Finder'' are used to select ``$\rm{Ap,b}$'' events. The condition is $\Delta\chi_{\rm 1L1S, 3L1S}^2>200$ and $\Delta\chi_{\rm \textcolor{cRsp2}{2Lb1S}, 3L1S}^2<200$. This means that there is planetary signal in the 3L1S light curve ($\Delta\chi_{\rm 1L1S, 3L1S}^2>200$), but the third mass $m_3$ is not required to explain the 3L1S light curve ($\Delta\chi_{\rm \textcolor{cRsp2}{2Lb1S}, 3L1S}^2<200$).} Similarly, we denote the event as ``$\rm{Ap,c}$'' if the detectable planet is $m_3$.
\item ``Abc'': Both planets are detectable in the 3L1S light curve. \textcolor{cRsp1}{Both ``Planet Finder'' and ``Triple-lens Finder'' are used to select these events. The condition is $\Delta\chi_{\rm 1L1S, 3L1S}^2>200$ and $\Delta\chi_{\rm \textcolor{cRsp2}{2L(best)1S}, 3L1S}^2>200$, where \textcolor{cRsp2}{$\rm 2L(best)1S$} is either \textcolor{cRsp2}{2Lb1S} or \textcolor{cRsp2}{2Lc1S}, depends on which one has smaller initial $\chi^2$ for fitting the 3L1S light curve.}
\end{itemize}

With the above definitions, we investigate the detectability of the Jupiter/Saturn analog (\S\ref{sec:res_det}) as well as the enhancement (\S\ref{sec:enh}) and suppression (\S\ref{sec:sup}) effects. We summarise the results in Tables \ref{tab:sum_prob}-\ref{tab:sum_5}.

\begin{table}
    \renewcommand\arraystretch{1.25}
    \centering
    \tabcolsep=0.02\columnwidth
    \caption{The detectability of two-planet events and the enhancement effect. \textcolor{mycolor1}{See \S\ref{sec:res_det} and \S\ref{sec:enh} for details.} }
    \begin{tabular}{c|cccc|c}
    \hline
    * & $\rm{Ab} \cap \rm{Ac}$ & $\rm{Ab} \cap \overline{\rm{Ac}}$ &  $\overline{\rm{Ab}} \cap \rm{Ac}$ & $\overline{\rm{Ab}} \cap \overline{\rm{Ac}}$ & Total\\
    \hline
    
    $P(\rm{*})$ & 1.29\% & 6.65\% & 1.94\% & 90.1\% & 100\% \\
    $P(\rm{Abc} \cap \rm{*})$ &1.10\% & 0.0387\% & 0.0299\% & $3.18\times 10^{-6}$ & 1.17\%\\
    \hline
    \end{tabular}
    \begin{tablenotes}
    \item Note: ``*'' has different meaning for different columns. For example, \textcolor{mycolor1}{for} the second column, ``*'' represents the event set \{$\rm{Ab}\cap\rm{Ac}$\}, so $P(*)$ represents $P(\rm{Ab}\cap\rm{Ac})$ and $P(\rm{Abc}\cap *)$ represents $P(\rm{Abc}\cap\rm{Ab}\cap\rm{Ac})$.
    \end{tablenotes}

    \label{tab:sum_prob}
\end{table}

\begin{table}
    \renewcommand\arraystretch{1.25}
    \centering
    \caption{The probabilities related to the suppression effect. \textcolor{mycolor1}{See \S\ref{sec:sup} for details.}   }
    \begin{tabular}{c|ccc|c}
    \hline
    * & $\rm{Ap,c}$ & $\rm{Ap,b}$ & $\overline{\rm{Ap}}$ & Total ($\overline{\rm{Abc}}$)\\
    \hline
    $P(* | (\rm{Ab} \cap \rm{Ac}))$ &1.72\% & 13.0\% & 0.119\% & 14.9\%\\
    \hline
    \end{tabular}
    \label{tab:sum_supp}
\end{table}

\begin{table}
    \renewcommand\arraystretch{1.25}
    \centering
    \caption{The detectability of both of the two planets given one of the planets is already detectable \textcolor{mycolor1}{(see \S\ref{sec:dep_u0} for details).}}
    \begin{tabular}{c|cc}
    \hline
    * & $\rm{Ab}$ & $\rm{Ac}$\\
    \hline
    
    $P(\rm{*})$ & 7.94\% & 3.23\% \\
    $P(\rm{Abc}|\rm{*})$ & 14.7\% & 36.1\% \\
    $P(\rm{Abc}| (\rm{*}\cap \log u_0\leq -1) )$ & 26.9\% & 67.2\% \\
    $P(\rm{Abc}| (\rm{*}\cap \log u_0\leq -2) )$ & 81.5\% & 83.8\% \\
    \hline
    \end{tabular}
    \label{tab:sum_3}
\end{table}

\begin{table}
    \renewcommand\arraystretch{1.25}
    \centering
    \caption{The dependence of the detectability of two-planet events on the impact factor \textcolor{mycolor1}{(see \S\ref{sec:dep_u0} for details).}}
    \begin{threeparttable}
    \begin{tabular}{c|ccc}
    \hline
    * & $\log u_0\leq -1$ & $\log u_0\leq -1.63$ & $\log u_0\leq -2$\\
    \hline

    $P(\rm{*})$ & 4.96\%\textcolor{cRsp1}{\tnote{a}} & 1.13\% & 0.454\% \\
    $P(\rm{Abc}|\rm{*})$ & 18.7\% & 50.2\% & 76.9\% \\
    
    \hline
    \end{tabular}
    
    \begin{tablenotes}
    \item[a] \textcolor{cRsp1}{We note that we draw $u_0$ between [0.001, 2] with the importance sampling method. Thus we have $P(\log u_0\leq -1)=(0.1-0.001)/(2-0.001)\sim4.95\%$ and $P(\log u_0\leq-2)=(0.01-0.001)/(2-0.001)\sim0.450\%$. These probabilities can also be calculated with Equation \eqref{equ:equprob}. }
    
    \end{tablenotes}
    \end{threeparttable}
    \label{tab:sum_4}
\end{table}

\begin{table}
    \renewcommand\arraystretch{1.25}
    \centering
    \caption{The dependence of the detectability of two-planet events on whether the two planets are inside the lensing zone $(0.6<s<1.6)$. \textcolor{mycolor1}{See \S\ref{sec:dep_sep} for details.}}
    \begin{tabular}{c|cccc}
    \hline
    * & $s_{2,\rm{in}} \cap s_{3,\rm{in}}$ & $s_{2,\rm{in}} \cap \overline{s_{3,\rm{in}}}$ &  $\overline{s_{2,\rm{in}}} \cap s_{3,\rm{in}}$ & $\overline{s_{2,\rm{in}}} \cap \overline{s_{3,\rm{in}}}$\\
    \hline
    $P(\rm{*})$ & 37.2\% & 40.9\% & 12.9\% & 8.99\% \\
    $P(\rm{Abc} | \rm{*})$ &2.35\% & 0.482\% & 0.623\% & 0.152\% \\
    \hline
    \end{tabular}
    \label{tab:sum_5}
\end{table}



\newcommand{\PPAbc}{1.17\%}
\newcommand{\PPAbcAbAc}{1.10\%}
\newcommand{\PPAbcAbNotAc}{0.0387\%} 
\newcommand{\PPAbcNotAbAc}{0.0299\%} 
\newcommand{\PPAbcNotAbNotAc}{3.18\times10^{-6}}
\newcommand{\PNotsup}{85\%} 

\subsection{Detectability of the scaled Sun-Jupiter-Saturn system}
\label{sec:res_det}

With the ``Planet Finder'' and ``Triple-lens Finder'', we can evaluate whether the two planets $m_2$ and $m_3$ are both detectable for a given pair of triple-lens system and source trajectory. By averaging the results from all simulated events, we obtain the overall detection probability $P(\rm{Abc})\sim \PPAbc$ \textcolor{mycolor0}{(the last column of Table \ref{tab:sum_prob})}. The set of events \{Abc\} can be divided into four \textcolor{mycolor1}{categories} \textcolor{mycolor0}{(see other columns of Table \ref{tab:sum_prob})},
\begin{itemize}
\item  ``Abc $\cap$ (Ab $\cap$ Ac)'', which means that $m_2$, $m_3$ are detectable from the corresponding 2L1S light curves with the ``Planet Finder'', and the two-planet signal \textcolor{cRsp1}{is} detectable in the 3L1S light curve with the ``Triple-lens Finder''. We find $P(\rm{Abc} \cap \rm{Ab} \cap \rm{Ac})=\PPAbcAbAc$, i.e., the majority \textcolor{mycolor1}{($1.10\%/1.17\% \approx 94\%$)} of events in set \{Abc\} are contributed by the set \{Ab $\cap$ Ac\}. 
\item  ``Abc $\cap$ (Ab $\cap$ $\overline{\rm {Ac}}$)'', which means that $m_3$ is undetectable from the 2L1S light curve corresponds to the lens system $m_1$-$m_3$ with the ``Planet Finder'', but it can be detected in the 3L1S light curve. We find $P(\rm{Abc} \cap \rm{Ab} \cap \overline{\rm{Ac}}) = \PPAbcAbNotAc$.
\item  ``Abc $\cap$ ($\overline{\rm {Ab}}$ $\cap$ Ac)'', which means that $m_2$ is undetectable from the 2L1S light curve corresponds to the lens system $m_1$-$m_2$ with the ``Planet Finder'', but it can be detected in the 3L1S light curve. We find $P(\rm{Abc} \cap \overline{\rm{Ab}} \cap \rm{Ac}) = \PPAbcNotAbAc$.
\item  ``Abc $\cap$ ($\overline{\rm {Ab}}$ $\cap$ $\overline{\rm {Ac}}$)'', which means that $m_2$, $m_3$ are undetectable from the corresponding 2L1S light curves with the ``Planet Finder''. However, the two-planet signal is detectable in the 3L1S light curve with the ``Triple-lens Finder''. \textcolor{mycolor1}{This category is very rare.} We find $P(\rm{Abc} \cap \overline{\rm{Ab}} \cap \overline{\rm{Ac}}) = \PPAbcNotAbNotAc$. 

\end{itemize}



Around $6\%$ of the detectable two-planet events are from the set \{$\overline{\rm{Ab} \cap \rm{Ac}}$\}, i.e., the detectability of two-planet events \textcolor{mycolor1}{is} enhanced \textcolor{mycolor1}{(see \S\ref{sec:enh} for more details)}. On the other hand, not all events in the set \{$\rm{Ab} \cap \rm{Ac}$\} contribute to the set \{Abc\}, i.e., the detectability of two-planet events \textcolor{mycolor1}{is} suppressed. We find $P(\rm{Abc} |( \rm{Ab} \cap \rm{Ac})) =\textcolor{mycolor1}{1.10\%/1.29\% \approx} \PNotsup$ \textcolor{mycolor1}{(see \S\ref{sec:sup} for more details)}.

\subsection{Enhancement effect}\label{sec:enh}

\begin{figure*}
    \centering
    \includegraphics[width=\textwidth]{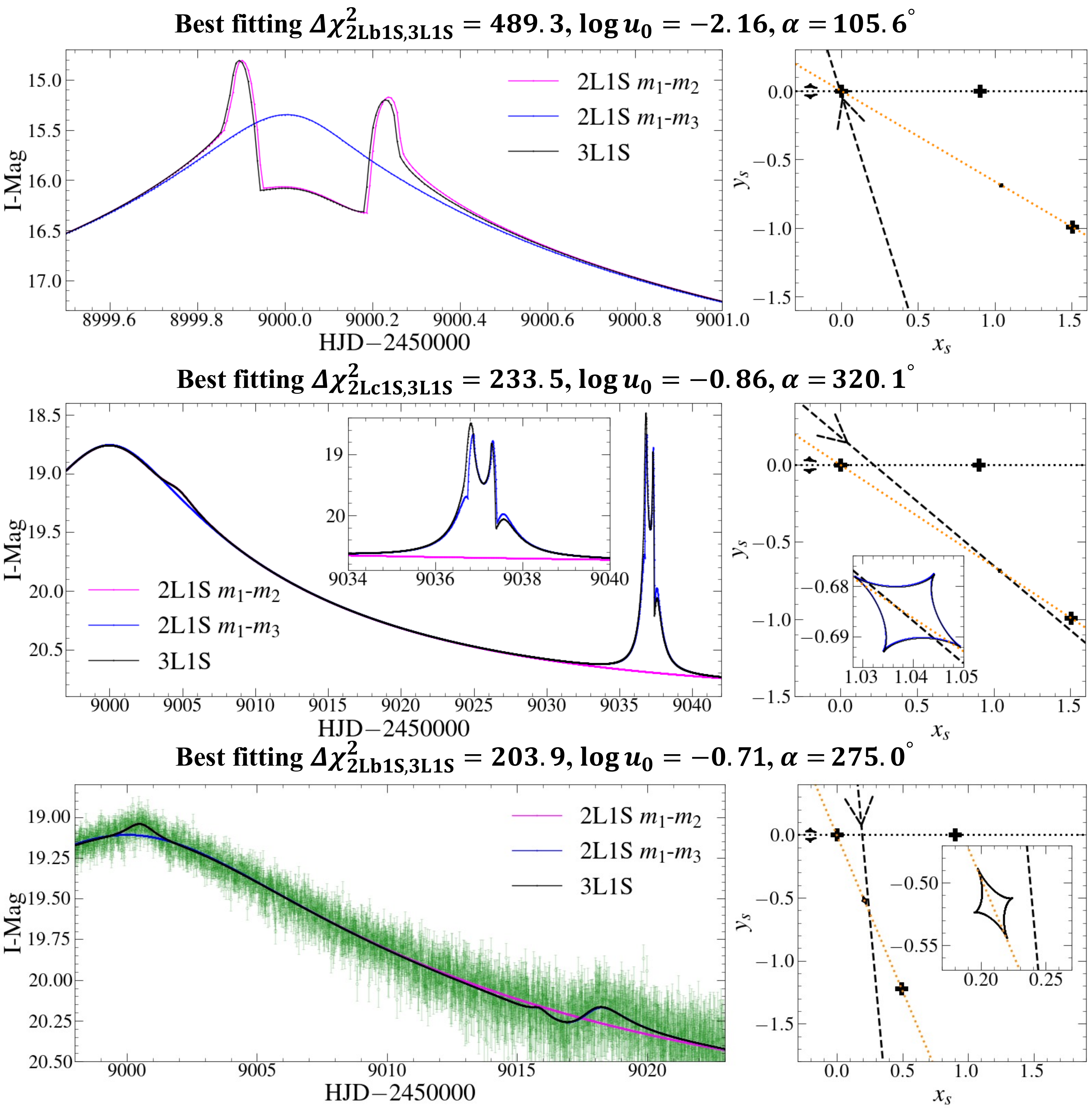}
    \caption{Example cases of the enhancement effect. The left panels show the model light curves, the right panels show the caustics and source trajectories. The upper panel corresponds to the set \{$\rm{Abc} \cap \rm{Ab} \cap \overline{\rm{Ac}}$\}, i.e., the detectability of $m_3$ is enhanced by the presence $m_2$. The middle panel corresponds to the set \{$\rm{Abc} \cap \overline{\rm{Ab}} \cap \rm{Ac}$\}, i.e., the detectability of $m_2$ is enhanced by the presence $m_3$. The bottom panel corresponds to the set \{$\rm{Abc} \cap \overline{\rm{Ab}} \cap \overline{\rm{Ac}}$\}, i.e., the detectability of $m_2$ and $m_3$ are mutually enhanced by each other\textcolor{cRsp1}{, the green points with error bars are the triple-lens light curve.} The solid curves with different colours show the light curves corresponding to different lenses. In each panel, we labelled the best-fitting $\Delta\chi^2$ of the ``Triple-lens Finder''. \textcolor{cRsp1}{In the inset of the middle right panel, we show the caustics (in blue) of the $m_1$-$m_3$ lens system on top of the caustics (in black) of the triple-lens system, the tiny difference in the caustics leads to the difference in the model light curves around $\rm{HJD}-2450000=9036$-9038.}} 
    \label{fig:enh}
\end{figure*}

Among the detectable triple-lens events, $\sim$94\% come from the set \{Ab $\cap$ Ac\}, i.e., $m_2$ and $m_3$ are detectable from the corresponding 2L1S light curves with the ``Planet Finder'' \textcolor{mycolor0}{(see Table \ref{tab:sum_prob}, $1.10/1.17=0.94$)}. Apart from this, about \textcolor{mycolor0}{$1-0.94 = 6\%$} of \{Abc\} come from the cases where $m_2$ or/and $m_3$ are undetectable in the 2L1S light curves. We regard these cases as the enhancement effect. Figure \ref{fig:enh} shows three example cases of the enhancement effect. The upper panel corresponds to the set \{$\rm{Abc} \cap \rm{Ab} \cap \overline{\rm{Ac}}$\}, where the detectability of $m_3$ is enhanced by the presence of $m_2$. For the 2L1S light curve with lenses $m_1$-$m_3$, the best-fitting 1L1S model in the ``Planet Finder'' has $\Delta\chi_{\rm{1L1S,\textcolor{cRsp2}{2Lc1S}}}^2=122.7$, i.e., the planet $m_3$ is undetectable. However, with the presence of $m_2$, the two-planet signal is detectable with $\Delta\chi_{\rm{\textcolor{cRsp2}{2Lb1S},3L1S}}^2=489.3$ after applying the ``Triple-lens Finder'' to the 3L1S light curve. This is 
due to the magnification pattern of the $m_1$-$m_2$ system \textcolor{mycolor1}{being} perturbed by $m_3$ in \textcolor{mycolor1}{the} region near the central caustics (see Figure \ref{fig:egmagmap}), so the signal of $m_3$ might be detectable under \textcolor{mycolor1}{favourable} source trajectories. 

Contrary to the previous case, the middle panel of Figure \ref{fig:enh} corresponds to the set \{$\rm{Abc} \cap \overline{\rm{Ab}} \cap \rm{Ac}$\}, where the detectability of $m_2$ is enhanced by the presence of $m_3$. The best-fitting 1L1S model for the 2L1S light curve with lenses $m_1$-$m_2$ has $\Delta\chi_{\rm{1L1S,\textcolor{cRsp2}{2Lb1S}}}^2=166.8$. With the presence of $m_3$, the two-planet signal is detectable with $\Delta\chi_{\rm{\textcolor{cRsp2}{2Lc1S},3L1S}}^2=233.5$. This is due to the magnification pattern in region around the planetary caustics of the $m_1$-$m_3$ system is perturbed by the presence of $m_2$ (see Figure \ref{fig:egmagmap_2}). 

Finally, the bottom panel of Figure \ref{fig:enh} corresponds to the set \{$\rm{Abc} \cap \overline{\rm{Ab}} \cap \overline{\rm{Ac}}$\}, where the detectability of $m_2$, $m_3$ are mutually enhanced. \textcolor{mycolor1}{Each of the two planets are undetectable individually with the ``Planet Finder'' in the corresponding 2L1S light curves, with $\Delta\chi_{\rm{1L1S,\textcolor{cRsp2}{2Lb1S}}}^2=143.5$ and $\Delta\chi_{\rm{1L1S,\textcolor{cRsp2}{2Lc1S}}}^2=168.4$ for $m_2$ and $m_3$, respectively. However, when applied to the 3L1S light curve, the ``Planet Finder'' has $\Delta\chi_{\rm{1L1S\textcolor{cRsp1}{,3L1S}}}^2=367.8$, and the ``Triple-lens Finder'' has $\Delta\chi_{\rm{\textcolor{cRsp2}{2Lb1S},3L1S}}^2=203.9$. \textcolor{cRsp1}{The two anomalies corresponding to individual planets are both present in the triple-lens light curve, which leads to the enhancement effect. The ``Triple-lens Finder'' uses the parameters of the 2L1S $m_1$-$m_2$ model parameters as the initial guess. This is because the anomaly caused by the $m_1$-$m_2$ system has higher magnification and smaller error bars in the data points (during $\rm{HJD}-2450000=8999$ - 9002) than the anomaly caused by $m_1$-$m_3$ system (during $\rm{HJD}-2450000=9014$ - 9022). If we take the 2L1S $m_1$-$m_3$ model parameters, the initial $\chi^2$ is larger by $\sim$110.}} 

We find that (see Table \ref{tab:sum_prob}):
\begin{itemize}
\item The detectability of $m_3$ is enhanced by the presence $m_2$: $P(\rm{Abc} |( \rm{Ab} \cap \overline{\rm{Ac}}))$ \textcolor{mycolor0}{$=0.0387\%/6.65\%$} $=0.582\%$, where $P(\rm{Ab} \cap \overline{\rm{Ac}})=6.65\%$ \textcolor{mycolor0}{(see the third column of Table \ref{tab:sum_prob})}.
\item The detectability of $m_2$ is enhanced by the presence $m_3$: $P(\rm{Abc} |( \overline{\rm{Ab}} \cap \rm{Ac}))=1.54\%$, where $P(\overline{\rm{Ab}} \cap \rm{Ac})=1.94\%$ \textcolor{mycolor0}{(see the fourth column of Table \ref{tab:sum_prob})}.
\item The detectabilities of $m_2$, $m_3$ are mutually enhanced: $P(\rm{Abc} | (\overline{\rm{Ab}} \cap \overline{\rm{Ac}}))=3.52\times10^{-6}$, where $P( \overline{\rm{Ab}} \cap \overline{\rm{Ac}})=90.1\%$ \textcolor{mycolor0}{(see the fifth column of Table \ref{tab:sum_prob})}.
\end{itemize}

\subsection{Suppression effect}\label{sec:sup}

\begin{figure*}
    \centering
    \includegraphics[width=\textwidth]{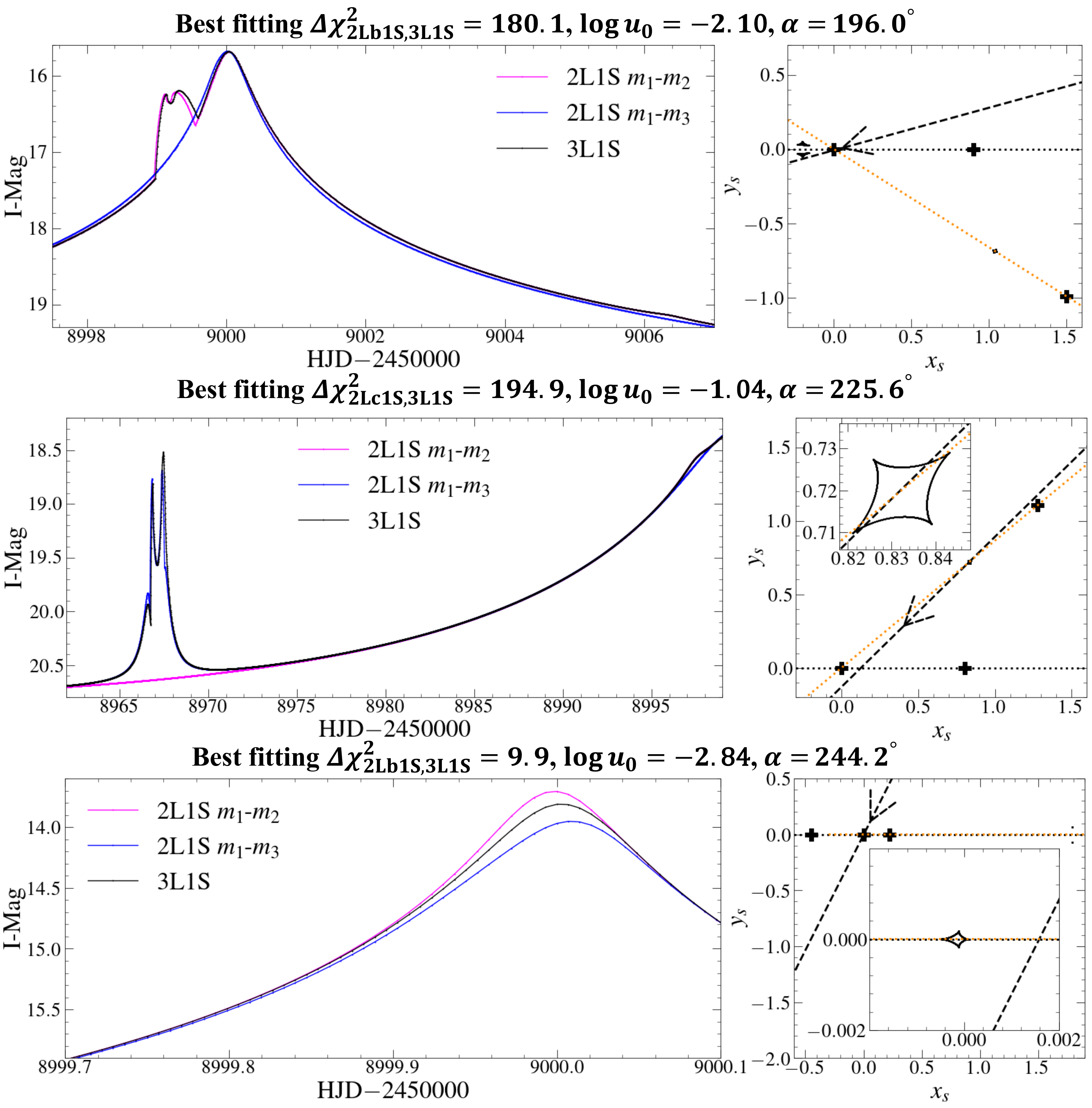}
    \caption{Similar to Figure \ref{fig:enh}, but for the suppression effect. The upper panel corresponds to the set \{$\rm{Ap,b} \cap \rm{Ab} \cap \rm{Ac}$\}, where the detectability of $m_3$ is suppressed by the presence of $m_2$. The middle panel corresponds to the set \{$\rm{Ap,c} \cap \rm{Ab} \cap \rm{Ac}$\}, where the detectability of $m_2$ is suppressed by the presence of $m_3$. The bottom panel corresponds to the set \{$\overline{\rm{Ap}} \cap \rm{Ab} \cap \rm{Ac}$\}, where the detectability of $m_2$ and $m_3$ are both suppressed.}
    \label{fig:sup}
\end{figure*}

Not all events in set \{$\rm{Ab} \cap \rm{Ac}$\} correspond to detectable two-planet events. Figure \ref{fig:sup} shows three example cases of the suppression effect. The upper panel corresponds to the set \{$\rm{Ap,b} \cap \rm{Ab} \cap \rm{Ac}$\}, where the detectability of $m_3$ is suppressed by the presence of $m_2$. The ``Planet Finder'' has $\Delta\chi_{\rm{1L1S,\textcolor{cRsp2}{2Lc1S}}}^2=306.2$ when finding the planet signal in the 2L1S light curve corresponding to lens $m_1$-$m_3$, i.e., $m_3$ is detectable with the absence of $m_2$. However, with the presence of $m_2$, the 3L1S light curve can be explained by a 2L1S model (with lenses $m_1$-$m_2$) with $\Delta\chi_{\rm{\textcolor{cRsp2}{2Lb1S},3L1S}}^2=180.1$, i.e., $m_3$ is no longer ``detectable'' in the 3L1S light curve, since \textcolor{cRsp1}{a 2L1S model with $m_1$-$m_2$ (\textcolor{cRsp2}{2Lb1S}) is already able to explain the 3L1S light curve where $\Delta\chi_{\rm{\textcolor{cRsp2}{2Lb1S},3L1S}}^2$}  falls below our threshold of 200.

The middle panel of Figure \ref{fig:sup} corresponds to the set \{$\rm{Ap,c} \cap \rm{Ab} \cap \rm{Ac}$\}, where the detectability of $m_2$ is suppressed by the presence of $m_3$. The best-fitting 1L1S model for the 2L1S light curve with lenses $m_1$-$m_2$ has $\Delta\chi_{\rm{1L1S,\textcolor{cRsp2}{2Lb1S}}}^2=235.9$\textcolor{cRsp2}{, because of anomaly at 8997.5, i.e., close to the primary peak}. However, with the presence of $m_3$, the two-planet signal is undetectable with $\Delta\chi_{\rm{\textcolor{cRsp2}{2Lc1S},3L1S}}^2 = 194.9$, the 3L1S light curve can be explained by a 2L1S model (with lenses $m_1$-$m_3$).

Finally, the bottom panel of Figure \ref{fig:sup} corresponds to the set \{$\overline{\rm{Ap}} \cap \rm{Ab} \cap \rm{Ac}$\}, where the detectability of $m_2$ and $m_3$ are both suppressed. Each of the two planets are detectable individually with the ``Planet Finder'' in the corresponding 2L1S light curves, with $\Delta\chi_{\rm{1L1S,\textcolor{cRsp2}{2Lb1S}}}^2=207.5$ and $\Delta\chi_{\rm{1L1S,\textcolor{cRsp2}{2Lc1S}}}^2=379.9$ for $m_2$ and $m_3$, respectively. However, when applied to the 3L1S light curve, the ``Planet Finder'' has $\Delta\chi_{\rm{1L1S\textcolor{cRsp1}{,3L1S}}}^2=190.1$\textcolor{mycolor1}{, below our threshold of 200}.

We summarize different probabilities as the following (see Table\textcolor{cRsp2}{s \ref{tab:sum_prob} and \ref{tab:sum_supp}}):
\begin{itemize}
\item Overall suppression \textcolor{mycolor1}{rate}: $P(\overline{\rm{Abc}} | (\rm{Ab} \cap \rm{Ac})) = 1-P(\rm{Abc} | (\rm{Ab} \cap \rm{Ac})) = 1 - 1.10\%/1.29\% \approx 15\%$ \textcolor{cRsp2}{(see the second column of Table \ref{tab:sum_prob})}.
\item The detectability of $m_2$ is suppressed by $m_3$: $P( \rm{Ap,c} | \textcolor{cRsp2}{(}\rm{Ab} \cap \rm{Ac}\textcolor{cRsp2}{)} )=1.72\%$ \textcolor{mycolor0}{(see the second column of Table \ref{tab:sum_supp})}.
\item The detectability of $m_3$ is suppressed by $m_2$: $P( \rm{Ap,b} | \textcolor{cRsp2}{(}\rm{Ab} \cap \rm{Ac}\textcolor{cRsp2}{)} )=13.0\%$ \textcolor{mycolor0}{(see the third column of Table \ref{tab:sum_supp})}. This means $m_3$ is ``overshadowed'' by $m_2$ more significantly.
\item The detectability of $m_2$ and $m_3$ is mutually suppressed by each other: $P( \overline{\rm{Ap}} | \textcolor{cRsp2}{(}\rm{Ab} \cap \rm{Ac}\textcolor{cRsp2}{)} )=0.119\%$ \textcolor{mycolor0}{(see the fourth column of Table \ref{tab:sum_supp})}.
\end{itemize}
The suppression effect is mainly contributed by cases where the detectability of $m_3$ is suppressed by the presence of $m_2$, i.e., the signal of $m_3$ is ``buried'' in the signal of $m_2$. We note that the suppression effect is related to the relative separations and mass ratios of the two planets. In our simulation, the intrinsic separations of $m_2$ and $m_3$ are fixed as $s_{\rm{2, true}}$ and $s_{\rm{3, true}}$, respectively. Among the 200 random projections, the ratio $s_3/s_2$ peaks at $\sim$2, and about half of all triple-lens systems have $s_3/s_2\in(1.7, 2.3)$. We leave the investigation of the detailed dependence of the suppression effect on the separations and mass ratios \textcolor{mycolor1}{to} a future work.

We note that \textcolor{mycolor1}{the} net effect of $m_2$ on the detectability of $m_3$ is suppression, i.e., $P( \rm{Ap,b} | \textcolor{cRsp2}{(}\rm{Ab} \cap \rm{Ac}\textcolor{cRsp2}{)} )\times P(\rm{Ab} \cap \rm{Ac}) - P(\rm{Abc} |( \rm{Ab} \cap \overline{\rm{Ac}})) \times P(\rm{Ab} \cap \overline{\rm{Ac}}) = 0.129\% > 0$. While $P(\rm{Ap,c} | \textcolor{cRsp2}{(}\rm{Ab} \cap \rm{Ac}\textcolor{cRsp2}{)} )\times P(\rm{Ab} \cap \rm{Ac}) - P(\rm{Abc} |( \overline{\rm{Ab}} \cap \rm{Ac})) \times P(\overline{\rm{Ab}} \cap \rm{Ac}) = -8\times10^{-5} \approx 0$, i.e., $m_3$ has a negligible net effect on the detectability of $m_2$.



\subsection{Dependence on parameters}

\subsubsection{Impact parameter}\label{sec:dep_u0}

\begin{figure}
    \centering
    \includegraphics[width=\columnwidth]{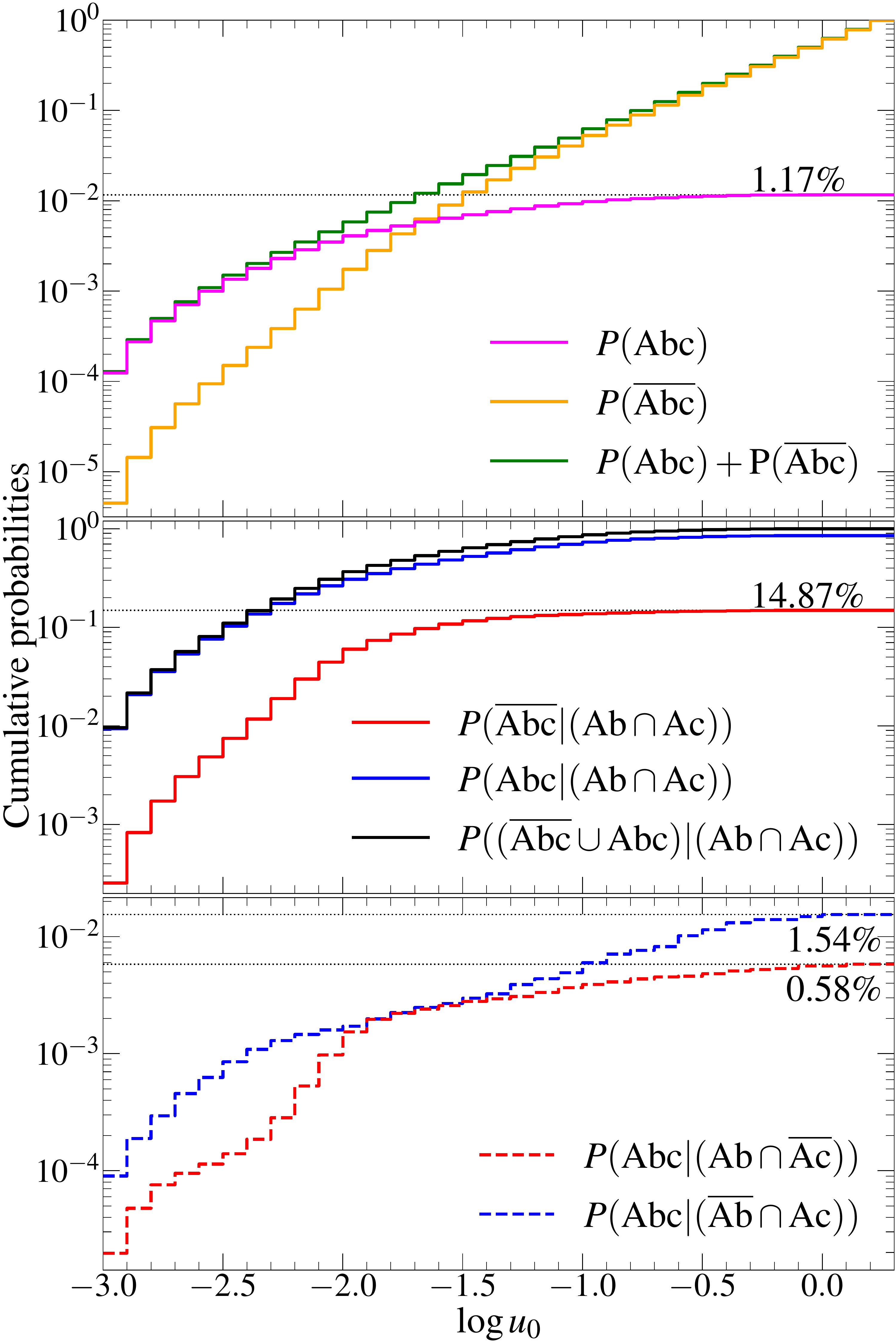}
    \caption{Cumulative probabilities as a function of $\log u_0$. The solid magenta line in the upper panel corresponds to the detectability of the ``Abc'' events. The solid red line in the middle panel corresponds to the probability of the suppression effect. In the lower panel, the dashed red line corresponds to the probability that the detectability of $m_3$ is enhanced by the presence of $m_2$, while the dashed blue line corresponds to the probability that the detectability of $m_2$ is enhanced by the presence of $m_3$.}
    \label{fig:prob_vs_u0}
\end{figure}

High magnification events are favoured in multiple-planet detections \citep[e.g.,][]{Griest1998, Gaudi1998ApJ, Zhu2014ApJ}. Among the five firmly established two-planet events listed in Table \ref{tab:alltriple}, four have $|u_0|\lesssim 10^{-2}$. Figure \ref{fig:prob_vs_u0} shows the cumulative probabilities as a function of $\log u_0$. The solid magenta line in the upper panel corresponds to the detectability of the ``Abc'' events. The solid red line in the middle panel corresponds to the probability of the suppression effect. In the lower panel, the dashed red line corresponds to the probability that the detectability of $m_3$ is enhanced by the presence of $m_2$, while the dashed blue line corresponds to the probability that the detectability of $m_2$ is enhanced by the presence of $m_3$. 

Both the detection probability of two-planet events and the suppression effect show strong dependence on the impact parameter. In the upper panel of Figure \ref{fig:prob_vs_u0} (see also Table \ref{tab:sum_4}), more than half of the events with $\log u_0$ smaller than $-1.63$ are detectable as two-planet events. The importance of high magnification events in detecting multiple-planet systems can be seen from another perspective (Table \ref{tab:sum_3}). The probability that $m_3$ is detectable given that the $m_2$ is detectable, i.e., $P(\rm{Abc}|\rm{Ab})$, is $14.7\%$ \textcolor{mycolor0}{(the third row of Table \ref{tab:sum_3})}. If we consider only high magnification events, e.g., for $\log u_0\leq-2$, this probability is $P(\rm{Abc}|\rm{Ab}, \log u_0\leq -2) = 81.5\%$. For comparison, $P(\rm{Abc}|\rm{Ac}) = 36.1\%$ and $P(\rm{Abc}|\rm{Ac}, \log u_0\leq -2) = 83.8\%$. \textcolor{mycolor1}{This implies} there is a high probability of detecting a second planet in a high magnification event if the first planet is detectable, which is consistent with the result of \cite{Gaudi1998ApJ}.

\begin{figure}
    \centering
    \includegraphics[width=\columnwidth]{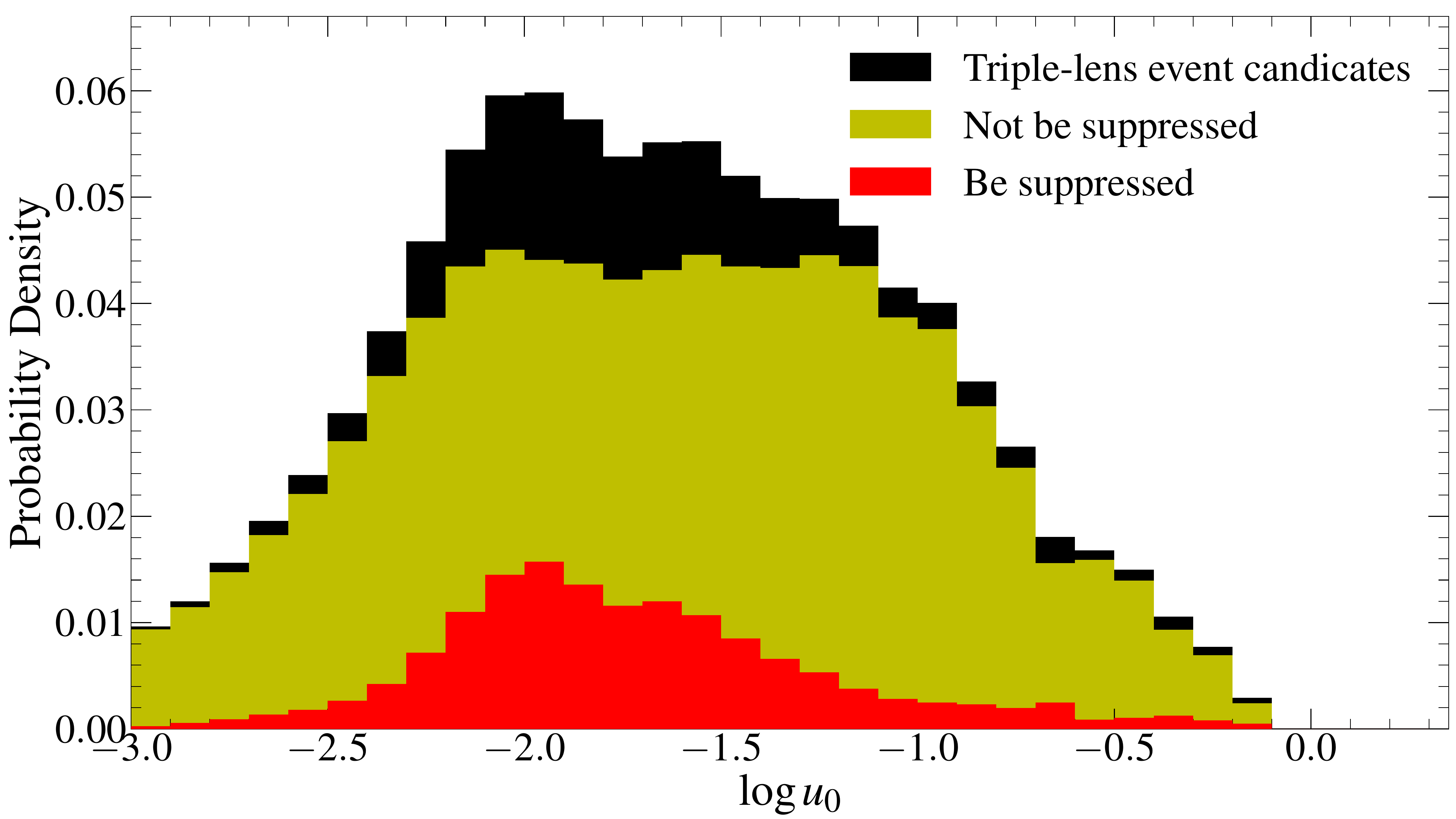}
    \caption{The dependence of the probability density for the suppression effect on the impact parameter. The black bars show the probability density of the triple-lens event candidates \{$\rm{Ab} \cap \rm{Ac}$\}. The red bars show the probability density of the suppressed candidates \{$\overline{\rm{Abc}}$\} among \{$\rm{Ab} \cap \rm{Ac}$\}.}
    \label{fig:prob_sup_eachbin}
\end{figure}

We show the dependence of the suppression probability density on the impact parameter in Figure \ref{fig:prob_sup_eachbin}. The black bars show the probability density of the triple-lens event candidates \{$\rm{Ab} \cap \rm{Ac}$\}. The red bars show the probability density of the suppressed candidates \{$\overline{\rm{Abc}}$\} among \{$\rm{Ab} \cap \rm{Ac}$\}. There is no simple relation between the suppression probability and the impact \textcolor{mycolor1}{parameter}. The suppression probability is small ($\lesssim 0.2\%$) when $\log u_0$ is small ($\lesssim -2.5$). This is reasonable since small impact \textcolor{cRsp1}{parameter} allows the source trajectory to interact with the central perturbations caused by both planets (e.g., see the right panel of Figure \ref{fig:egmagmap}). So the planet signatures are hard to be suppressed. The suppression probability peaks at $\log u_0 \approx -2$. A possible reason is that at such $u_0$, the signal of the Jovian planet is still strong, but the signal of the Saturn-like planet is not strong enough\textcolor{mycolor1}{, so it can} be buried more easily (see the upper panel of Figure \ref{fig:sup}). \textcolor{cRsp1}{For} $\log u_0 \gtrsim -2$, the suppression probability decreases as $u_0$ increases. The reason \textcolor{mycolor1}{may be} that \textcolor{mycolor1}{a} larger $u_0$ allows the source trajectory to interact with the planetary caustic\textcolor{mycolor1}{s} correspond\textcolor{mycolor1}{ing} to the Saturn-like planet.



\subsubsection{Planet-host separation}\label{sec:dep_sep}


For about $37.2\%$\footnote{Given two planets with intrinsic true separations $s_{2,\rm{true}} = 0.903$ and $s_{3,\rm{true}} = 1.801$, with random orbital inclination and orbital phases, the theoretical probability that they have projected separations both inside the standard lensing zone is $34\%$. In this work we obtain a different value (37\%) due to the numerical noise caused by a limited number (200) of projection realisations.} of our simulated events, the two planets are simultaneously projected inside the standard lensing zone \citep[$0.6<s_2, s_3<1.6$,][]{Griest1998}, among them $2.35\%$ are detected as two-planet events. In comparison, only $0.464\%$ of the events where the two planets are not simultaneously inside the standard lensing zone are detected as two-planet events, indicat\textcolor{mycolor1}{ing} multiple planets are more likely to be \textcolor{mycolor1}{simultaneously detected if all of them are located} inside the lensing zone (Table \ref{tab:sum_5}).

We note that among the five firmly established two-planet microlensing events as listed in Table \ref{tab:alltriple}, almost all planets are located inside the lensing zone, except for OGLE-2018-BLG-1011, where the larger planet has separation $s_2=0.582$, \textcolor{mycolor1}{just outside the inner lensing zone (0.6)}.




\subsubsection{\textcolor{mycolor1}{$\Delta\chi^2$} detection threshold}\label{sec:dep_on_dchisq}

\begin{figure}
    \centering
    \includegraphics[width=\columnwidth]{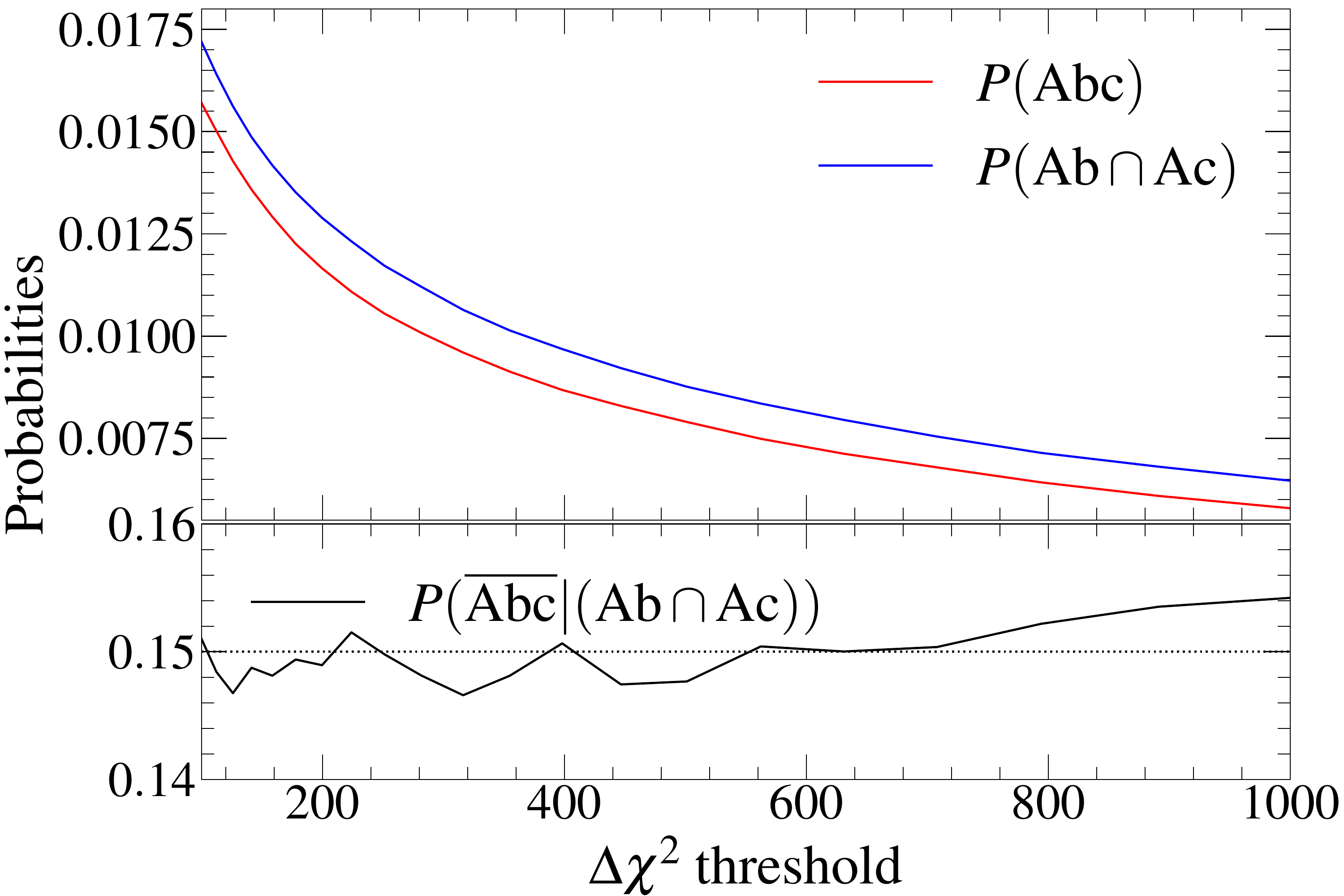}
    \caption{The detection probability of two-planet event candidates (\{$\rm{Ab} \cap \rm{Ac}$\}) and two-planet events (\{$\rm{Abc}$\}) as a function of $\Delta\chi^2$ threshold of the ``Planet Finder'' and ``Triple-lens Finder''. The lower panel shows the level of the suppression effect.}
    \label{fig:res_vs_dchisq}
\end{figure}

To investigate how the chosen $\Delta\chi^2$ threshold of the ``Planet Finder'' and ``Triple-lens Finder'' will influence the results, we re-calculate probabilities by using different $\Delta\chi^2$ thresholds. Figure \ref{fig:res_vs_dchisq} shows the result. \textcolor{mycolor1}{As expected, the} detection probability of both two-planet event candidates (\{$\rm{Ab} \cap \rm{Ac}$\}) and two-planet events (\{$\rm{Abc}$\}) both decrease as the $\Delta\chi^2$ threshold increase\textcolor{mycolor1}{s}. However, the suppression probability ($\sim$15\%) is not greatly affected. 

%% file: dis.tex

The microlensing magnification excess caused by planets makes them detectable \citep{Andy1992, Chung2005}. The pattern of the magnification excess map (e.g., Figures \ref{fig:egmagmap} and \ref{fig:egmagmap_2}) is complex, especially when there are multiple planets in the same system (\citealt{Gaudi1998ApJ, 2005Han}, see also \citealt{Danek2015, Danek2019} for the complex caustics caused by triple lenses). When there are multiple planets, the magnification excess caused by each planet \textcolor{mycolor1}{influence} each other, leading to the enhancement or suppression of their detectability. \textcolor{mycolor1}{Furthermore}, the magnification excess pattern changes as the projected positions of the lenses change, making it hard to evaluate the detection probability of a \textcolor{mycolor1}{given} triple-lens system \textcolor{mycolor1}{under different} instrumentation\textcolor{mycolor1}{s} and observing strateg\textcolor{mycolor1}{ies. In this work, we} simulated a large number of events \textcolor{cRsp1}{with cadence $\Gamma = 6\,\mathrm{hr}^{-1}$} to assess the detection probability of scaled Sun-Jupiter-Saturn system\textcolor{mycolor1}{s} with the \textcolor{cRsp1}{\tele}. We discuss the implications of our results \textcolor{mycolor1}{in} \S\ref{sec:implications}, and future works \textcolor{mycolor1}{in} \S\ref{sec:future}.

\subsection{The implications of \textcolor{mycolor0}{our} results}\label{sec:implications}

A few previous studies estimated the occurrence rate of multiplanetary systems in the microlensing field. \cite{mufun} conducted a detailed sensitivity analysis of 13 $\mu$FUN planetary events with magnification larger than 200, including a two-planet event \citep[OGLE-2006-BLG-109,][]{OB06109}. They investigated that if every microlensed star possessed a ``scaled version'' of the solar system, they would have detected 6.1 systems with two or more planet detections. They thus estimated that the frequency of solar-like systems is 1/6. \cite{OB141722} estimated \textcolor{mycolor1}{that} the occurrence rate of systems with two cold gas giants is $6\%\pm2\%$ based on two triple-lens events (OGLE-2006-BLG-109 and OGLE-2014-BLG-1722) and assuming that the detection efficiency of two-planet systems can be approximated as the product of detection efficiencies for each planet.

The suppression probability ($15\%$) found in this work has important implications for the calculations of detection efficiency and occurrence rate of microlensing multi-planetary systems. The upcoming microlensing surveys are expected to discover $\sim$100 two-planet systems. The occurrence rate estimation of two-planet systems based on these discoveries would have $\sim$10\% Poisson noise, which \textcolor{mycolor1}{may} be at the same level or even \textcolor{mycolor1}{below} the suppression probability. If we assume the detection efficiency of a two-planet system can be approximated as $\hat{\epsilon}_{\rm{Abc}}=\epsilon_{\rm{Ab}}\times\epsilon_{\rm{Ac}}$, the estimated value $\hat{\epsilon}_{\rm{Abc}}$ could deviate substantially from the real value. 


We note that there is no microlensing study on the estimation occurrence rate of planets in binary systems. A recent study noticed that all planets in binary systems published by the KMTNet survey are located inside the resonant caustics with mass ratio $\gtrsim 2 \times 10^{-3}$ \citep{OB191470}, implying the incompleteness of the KMTNet sample for planets in binary systems, or the suppression effect for the planet \textcolor{mycolor1}{may be more severe}. More detailed studies are \textcolor{mycolor1}{needed} to gain a better understanding.

\subsection{Future works}\label{sec:future}

The parameter space covered in this work is small. The known triple microlensing events (Table \ref{tab:alltriple}) and the results from exoplanet demographic surveys have shown the diversity of planetary systems \citep{Gaudi2021book}. \textcolor{mycolor1}{Further improvements can be made} in several ways.

The first is to investigate the detectability properties across a wider range of parameter space related to the lens, i.e., $(s_2, q_2, s_3, q_3$), \textcolor{mycolor1}{as} upcoming surveys will be sensitive to a broad range of masses and semi-major axes. \textcolor{mycolor1}{For example, on} the low mass end, the Roman telescope has the sensitivity to the mass of Ganymede \citep{MatthewWFIRSTI}\textcolor{mycolor1}{; s}everal works have investigated the detectability of extrasolar moons with the microlensing method \citep{2002HanHan, Liebig2010Exomoon, 2022EuclidRoman}.

The high mass end is also interesting \textcolor{mycolor1}{to explore}. One can investigate the detectability of planets in binary systems \textcolor{cRsp1}{(e.g., \citealt{Luhn2016ulensPtype, Han2017CR})} and how the binary star system will influence the detectability of the planet. It is known that stellar binarity affects planet formation and evolution \citep[e.g.,][]{Moe2021}. For example, most circumbinary planets detected by \textcolor{cRsp1}{the Kepler satellite} are located near the stability limit such that they would be dynamically unstable if they were in a slightly closer orbit \citep{1999HolmanStable, Ballantyne2021}. A possible explanation is that these planets formed further out, then migrated to the cavity edge (close to the location of the stability limit) truncated by the binary \citep{Kley2014}. However, \cite{Quarles2018} conducted $N$-body simulations and did not find strong evidence for a pile-up in the circumbinary planet systems \textcolor{cRsp1}{detected by the Kepler satellite}. \textcolor{mycolor0}{In addition, \cite{OB07349}} reported the first microlensing circumbinary planet with an orbit well beyond the stability limit. To test the degree to which stellar binarity inhibits or promotes planet formation, one needs more samples as well as a better understanding of the detectability properties, e.g., how the binary stars suppress or enhance the detectability of planets. \cite{Eggl2013ApJ} estimated the detectability of a terrestrial planet in coplanar S-type binary configurations by using the radial velocity, astrometry, and transit photometry methods. They found that the gravitational interactions between the second star and the planet can facilitate the planet's detection. There is no such study for the microlensing method.

The second is to carefully take into account the properties of the stellar populations in the galactic disk and bulge and their dynamics, which influence the related parameters such as $\te$, $\rho$, \textcolor{mycolor1}{$I_{\rm{blend}}$}, and $I_{\rm{source}}$. Although larger source radii increase the probability that the sources interact with the caustic structure, the finite-source effect will eliminate small deviations caused by low-mass objects. \cite{2002HanHan} found that detecting signals from satellites will be very difficult due to the signals being seriously smeared out by the finite-source effect. \cite{Gaudi2000ApJ} found that \textcolor{mycolor1}{the} finite-source effect can significantly influence the detection efficiency for an object with mass ratio $\lesssim 10^{-3}$. In addition, they found that the fraction of blended light is an important factor such that higher blend fractions imply \textcolor{mycolor1}{a} smaller $u_0$ \textcolor{mycolor1}{is needed} for detection.

The third is to investigate the detectability properties with different instruments and observing strategies. \cite{Shvartzvald_Maoz2012} quantified the dependence of detected planet yield on observation cadence and duration and found that the yield is doubled when the baseline cadence increases from $1/3\;\mathrm{hr}^{-1}$  to $4\;\mathrm{hr}^{-1}$, or when the \textcolor{mycolor1}{observation} duration increases from 80 days to 150 days. The upcoming microlensing surveys with different instruments and observing modes will lead to different yields. In addition, space $+$ space based \citep[e.g., Roman $+$ Euclid,][]{BacheletPenny2019ApJ, 2022EuclidRoman}, and space $+$ ground based \textcolor{mycolor0}{\citep[e.g., ET + KMTNet,][]{Gould2021RAA, ET2}} joint microlensing surveys have been proposed to better constrain the properties of microlensing events. 

Finally, the degeneracy in triple microlensing is more complex and needs to be considered carefully \textcolor{mycolor1}{\citep[e.g.,][]{Song2014}}. The \textcolor{mycolor1}{larger} parameter space \textcolor{mycolor1}{will} result in many pairs of degenerate triple-lens solutions in reality, especially when the data coverage is not dense enough. On the other hand, the triple-lens model can be degenerate with other models, such as the binary-lens-binary-source \textcolor{mycolor1}{(2L2S)} model, as demonstrated in several events (shown as triple-lens event candidates in Table \ref{tab:alltriple}). \textcolor{mycolor0}{In addition}, it would be more complicated if high-order effects are included, such as the microlens parallax effect \citep{Gould2000}, xallarap effect \citep{Poindexter2005}, and the lens orbital motion \citep{MB09387, OB09020, Penny2011Orbital}. The suppression effect might be more severe when 2L1S $+$ high-order effects are considered to explain 3L1S light curves. For example, MACHO-97-BLG-41 \citep{Bennett1999} and OGLE-2013-BLG-0723 \citep{OB130723} were once interpreted as 3L1S events, but later studies found 2L1S model with lens orbital motion is also feasible \citep{MACHO9741Albrow, MACHO9741Jung, OB130723Han}. A recent event shows severe degeneracy between 2L1S $+$ high-order effects and 3L1S interpretations, which implies that the degeneracy can be \textcolor{mycolor1}{quite} common \citep[KMT-2021-BLG-0322,][]{KB210322}. \textcolor{cRsp1}{Therefore, it may be necessary to explore the results with $\Delta \chi_{\rm{1L1S,2L1S}}^2 \neq \Delta \chi_{\rm{2L1S,3L1S}}^2.$}


\begin{table*}[t]
    \renewcommand\arraystretch{1.25}
    \centering
    \caption{List of triple-lens events.}
    \begin{threeparttable}
    \begin{tabular}{l|ccccccc}
    \hline
    Event name & $q_2 (10^{-3})$ & $s_2$ & $q_3 (10^{-3})$ & $s_3$ & $|\psi| (^\circ)$ & $|u_0| (10^{-3})$ & $\te$ (days)\\
    \hline
             &     \multicolumn{7}{c}{Firmly established two-planet systems}     \\
    OGLE-2006-BLG-109 \citep{OB06109, OB06109_Dave} & 1.36 & 0.627 & 0.506 & 1.04 & $13.5$ & 3.48 & 127\\
    OGLE-2012-BLG-0026 \citep{OB120026} & 0.784 & 1.25 & 0.130 & 1.03 & $137$ & 9.20 & 93.9 \\
    OGLE-2018-BLG-1011 \citep{OB181011} & 15.0 & 0.582 & 9.84 & 1.28 & $82.0$ & 53.0 & 12.4\\
    OGLE-2019-BLG-0468 \citep{OB190468} & 10.6 & 0.717 & 3.54 & 0.853 & 48.0 & 12.0 & 75.2 \\
    KMT-2021-BLG-1077 \citep{KB211077} & 1.75 & 0.973 & 1.56 & 1.31 & 31.5 & $10.6$ & 24.9\\
    \hline
    &     \multicolumn{7}{c}{Firmly established planets in binary systems}     \\
    OGLE-2006-BLG-284 \citep{OB06284} & 284 & 0.798 & 1.16 & 0.764 & $177$ & 77.5 & 39.7 \\
    OGLE-2007-BLG-349 \citep{OB07349} & 869 & 0.020 & 0.638 & 0.815 & 21.2 & 1.98 & 118\\
    OGLE-2008-BLG-092 \citep{OB08092} & 220 & 17.0 & 0.241 & 5.26 & 158 & 1545 & 38.6\\
    OGLE-2013-BLG-0341 \citep{OB130341} & 1211 & 12.9 & 0.0480 & 0.814 & 170 & 23.3 & 33.4 \\
    OGLE-2016-BLG-0613 \citep{OB160613} & 29.0 & 1.40 & 3.27 & 1.17 & $54.5$ & 21.0 & 74.6\\
    OGLE-2018-BLG-1700 \citep{OB181700} & 297 & 0.274 & 10.0 & 1.18 & 37.7 & 6.70 & 41.9 \\
    KMT-2019-BLG-1715 \citep{KB191715} & 246 & 0.551 & 4.01 & 1.05 & 77.2 & 56.0 & 44.0\\
    KMT-2020-BLG-0414 \citep{KB200414} & 57.8 & 0.0940 & 0.0113 & 0.999 & 41.9 & 0.688 & 94.4\\
    \hline
    &     \multicolumn{7}{c}{Triple-lens event candidates}     \\
    OGLE-2014-BLG-1722 \citep[Two-planets,][]{OB141722} & 0.639 & 0.851 & 0.447 & 0.754 & 126 & 131 & 23.8 \\
    OGLE-2018-BLG-0532 \citep[Two-planets,][]{OB180532} & 3.08 & 0.364 & 0.0975 & 1.01 & 2.06 & 8.23 & 139\\
    KMT-2019-BLG-1953 \citep[Two-planets,][]{KB191953} & 8.65 & 4.92 & 1.91 & 2.30 & 119 & 2.36 & 16.2\\
    OGLE-2019-BLG-0304 \citep[Planet in binary,][]{OB190304} & 1.82 & 0.885 & 0.363 & 3.73 & 138 & 0.540 & 17.8\\
    OGLE-2019-BLG-1470 \citep[Planet in binary,][]{OB191470} & 359 & 0.439 & 3.47 & 1.11 & 73.9 & 187 & 42.6 \\
    KMT-2021-BLG-0240 \citep[Two-planets,][]{KB210240} & 1.83 & 2.72 & 0.640 & 0.954 & 97.5 & 3.03 & 42.3\\
    \hline
    \end{tabular}
    \begin{tablenotes}
    
    \footnotesize
    \item Note: We show the solution with smallest $\chi^2$ when there are multiple degenerate triple-lens solutions. We unify the coordinate system used by different authors according to Equation (\ref{equ:coord}) and sort the masses as $q_2 > q_3$.
    \end{tablenotes}
    \label{tab:alltriple}
    \end{threeparttable}
\end{table*}

%% file: conclu.tex
The upcoming microlensing surveys are expected to discover a larger sample of triple microlensing events. We investigate the detectability of a scaled Sun-Jupiter-Saturn system within the context of \textcolor{cRsp1}{the \tele}. The key findings of our work are as follows,
\begin{enumerate}
\item The probability that \textcolor{mycolor0}{a} scaled Sun-Jupiter-Saturn system \textcolor{mycolor1}{being} detectable with the \textcolor{cRsp1}{\tele} is about $1\%$.
\item The presence of \textcolor{mycolor0}{a Saturn-like} planet has a negligible effect on the detection probability of the Jovian planet. While the presence of the Jovian planet suppresses the detectability of the Saturn-like planet by about $13\%$. This level of suppression is nearly constant regardless of the adopted $\Delta\chi^2$ threshold and is about the same order as the result \textcolor{mycolor0}{reported} in \cite{Zhu2014ApJ}, although their lenses are different \textcolor{mycolor0}{from} ours, indicating that the suppression effect is prevalent and \textcolor{mycolor0}{non-}negligible. Future statistical works on the sensitivity of multi-planetary microlensing events should be cautious in this regard. The assumption that the detection efficiency of multi-planetary systems is approximately the product of detection efficiency for each planet could lead to substantial errors.
\item There is no simple relation between the suppression probability and the impact \textcolor{cRsp1}{parameter}. The suppression probability peaks at $u_0\approx 0.01$ in the present work. For events with $-2.5\lesssim \log u_0 \lesssim -1.0$, it is not safe to assume that the detection efficiency of multiple planets can be calculated by treating each of the planets as \textcolor{mycolor0}{being} independent. 
\item High magnification events are important in discovering two-planet events. For the scaled Sun-Jupiter-Saturn lens \textcolor{mycolor0}{considered here}, more than half of the events with $\log u_0\leq-1.63$ are detectable as two-planet events with the \textcolor{cRsp1}{\tele}. The probability that $m_3$ \textcolor{mycolor0}{(Saturn-like)} is detectable given that $m_2$ \textcolor{mycolor0}{(Jovian)} is detectable, i.e., $P(\rm{Abc}|\rm{Ab})$, is $14.7\%$. While for high magnification events, e.g., \textcolor{mycolor0}{for} $\log u_0\leq-2$, this probability is $P(\rm{Abc}|\rm{Ab}, \log u_0\leq -2) = 81.5\%$. 
In turn, $P(\rm{Abc}|\rm{Ac}) = 36.1\%$ and $P(\rm{Abc}|\rm{Ac}, \log u_0\leq -2) = 83.8\%$.
\item If two planets are simultaneously located inside the lensing zone \textcolor{mycolor0}{($0.6<s<1.6$)}, the probability of discovering triple-lens events is higher. For about $37\%$ of our simulated events, the two planets are simultaneously inside the lensing zone, among \textcolor{mycolor0}{these} $2.4\%$ are detected as two-planet events. However, if the two planets are not simultaneously inside the lensing zone, this probability (0.46\%) becomes five times smaller. 
\end{enumerate}


